\documentclass[english,tightenlines,eqsecnum,amsmath,amssymb,aps,prd,superscriptaddress,nofootinbib,showpacs,floats]{article}
\usepackage{amsmath,amsfonts,amssymb,multicol}
\usepackage{graphicx,caption,subcaption}
\usepackage[usenames]{xcolor}
\definecolor{AHZ}{rgb}{0.0,1,0.0}
\usepackage[colorlinks,linkcolor=AHZ,citecolor=red,bookmarks]{hyperref}
\usepackage{rotating}
\usepackage[mathscr]{eucal}
\usepackage{graphicx}
\usepackage[T1]{fontenc}
\usepackage{babel}
\usepackage{authblk}
\usepackage{epsfig}
\usepackage{color}
\usepackage{amssymb}
\usepackage{fancyhdr}

\newcommand{\f}[2]{\frac{#1}{#2}}
\def\be{\begin{equation}}
	\def\ee{\end{equation}}
\def\bea{\begin{eqnarray}}
	\def\eea{\end{eqnarray}}

\def\bwt{\begin{widetext}}
	\def\ewt{\end{widetext}}

\usepackage{amsmath}

\usepackage{amssymb}

\begin{document}
\title{Dynamical  wormholes in {Lovelock} gravity}
\author{Mohammad Reza Mehdizadeh$^{1}$\thanks{mehdizadeh.mr@uk.ac.ir}}
\author{Amir Hadi Ziaie$^{2}$\thanks{ah.ziaie@maragheh.ac.ir}}
\affil{{\rm $^1$Department~of~ Physics,~ Shahid~ Bahonar~ University, P.~ O.~ Box~ 76175, Kerman, Iran}}
\affil{{\rm $^2$Research~Institute~for~Astronomy~and~Astrophysics~of~ Maragha~(RIAAM), University of Maragheh,  P.~O.~Box~55136-553,~Maragheh, Iran}}
\renewcommand\Authands{ and }
\maketitle
\begin{abstract}
In this study we present evolving wormhole configurations in third-order {Lovelock} gravity and investigate the possibility that these solutions satisfy the energy conditions. Using a generalized Friedmann-Robertson-Walker spacetime, we derive evolving wormhole geometries by considering a constraint on the Ricci scalar. In standard cosmological models, the Ricci scalar is independent of the radial coordinate $r$ and is only a function of time. We use this property to introduce dynamic wormhole solutions expanding in an inflationary cosmological background and explore the effects of higher-order {Lovelock} terms on the dynamics of such wormholes. Our analysis shows that for suitable third-order {Lovelock} coefficients, there are wormhole solutions that respect the weak energy condition {(WEC)}. In addition to this, we also present other wormhole solutions that satisfy the WEC throughout their respective evolution.
\end{abstract}
\section{Introduction}
Wormholes are throatlike geometrical structures which have a characteristic to connect two parallel Universes or otherwise distant parts of
the same Universe. The first study on the concept of wormholes was done by Ludwing Flamm in 1916~\cite{Flamm1916} as a \lq{}\lq{}tunnel structure\rq{}\rq{} in the Schwarzschild spacetime, however, after some time it was realized that the solutions he obtained are not relativistically applicable. After this work, a similar geometrical configuration was proposed by Einstein and Rosen~\cite{ERB}, a \lq{}\lq{}bridge structure\rq{}\rq{} between black holes in order to obtain a regular solution without a spacetime singularity. This solution is known as Einstein-Rosen bridge. The concept of traversable wormholes was firstly defined by Morris and Thorne in 1988~\cite{mt}. In their influential work, Morris and Thorne investigated static spherically symmetric wormhole configurations using the principles of general relativity (GR) and presented the fundamental theory for traversable wormholes. In this manner, it was found that these configurations possess a specific trait, i.e., the {\it exoticity} of supporting material. In other words, a traversable wormhole must obey the fundamental flaring-out condition at the wormhole throat. However, the fulfillment of this condition requires an unknown source of exotic matter so that the energy-momentum tensor originated from such types of matter violates the null energy condition (NEC)~\cite{khu,Visser1995}. 

As the classical forms of ordinary matter are believed to satisfy the standard energy conditions, then, one of the most important challenges in wormhole scenarios is the establishment of these conditions. Hence, due to its problematic nature, several avenues of research have been pursued in recent decades in order to minimize the usage of exotic matter. For instance, in~\cite{karvise}, a particular class of spacetime geometries admitting a traversable wormhole has been constructed using arbitrarily small quantities of exotic matter. In~\cite{Parsa2020}, it is shown that considering a general equation of state for the fluid profiles can act in favor of minimal usage of exotic matter. However, generalizing the matter sector may not necessarily guarantee obtaining a setting for nonexotic wormhole geometries~\cite{revworm}. An interesting method for creating wormhole structures was proposed by Visser and Poisson in~\cite{pv} where, using the cut and paste technique they confined the presence of exotic matter to a thin shell. Such an attractive approach led to the introduction of the so-called thin shell wormholes; different and interesting aspects of such objects have been studied extensively in the literature~\cite{thin1}. Like its counterpart in GR, the issue of exotic matter distribution in wormhole configurations has also been investigated in modified gravity theories, since the modifications to the GR theory may provide extra terms that support the wormhole geometry without resorting to an exotic matter source. In this respect, a huge amount of work has been carried out during the past years to build and study wormhole solutions within the framework of modified gravity theories; for example, the study of wormhole solutions in the framework of gravitational decoupling~\cite{Ortiz2021}, wormhole solutions in Brans-Dicke theory~\cite{bd}, $f(R)$ gravity~\cite{fr}, Born-Infeld theory~\cite{bf}, Einstein-Gauss-Bonnet gravity~\cite{bhakar,EGBWORM}, gravity theories with torsion~\cite{torsion}, Kaluza-Klein gravity~\cite{kl}, $f(Q)$ gravity with $Q$ being the nonmetricity term~\cite{WormQ}, and scalar-tensor gravity~\cite{kash}, and solutions in the presence of a cosmological constant have been reported in~\cite{anac}. Recently, the possibility of the existence of traversable	wormholes in the context of $f(R,T)$ gravity have been reported in the literature, such as wormhole formation with two types of varying Chaplygin gas~\cite{elmkh} and wormhole solutions satisfying the energy conditions in the exponential $f(R,T)$ gravity~\cite{mosa}; see also~\cite{frtt} for other solutions in this theory. 
\par
Though in the GR framework, static wormhole configurations require fluid sources that violate the NEC, there are nonstatic (evolving) Lorentzian wormholes without the need of exotic matter to sustain them. One way to investigate such objects is to embed a wormhole spacetime in a Friedmann-Lema{\rm$\hat{i}$}tre-Robertson-Walker (FLRW) metric, thus allowing the geometry to evolve in a cosmological background. It is therefore shown that the resulting wormhole structures have different characteristics compared to the static ones, namely, they can live for arbitrarily small or large time intervals~\cite{kar14} or even satisfy the dominant energy condition in the whole spacetime~\cite{DEC1516}. Further studies along this subject have been performed towards constructing dynamical wormhole geometries which satisfy the energy conditions and also the averaged energy conditions over timelike or null geodesics during a time period~\cite{dyamic1,dynamic01}. An interesting scenario is that the expansion of the Universe could increase the size of the static wormholes by a factor which is proportional to the scale factor of the Universe, in a time-dependent inflationary background~\cite{rom1}. In this respect, evolving wormholes in a cosmological background have been studied in~\cite{karqunt}, and dynamic wormhole spacetimes supported by two fluids and also by a polytropic equation of state have been investigated in~\cite{cata1} and~\cite{cata2}, respectively. Such configurations that respect energy conditions have been also reported in the framework of Einstein-Cartan gravity~\cite{mehdiziai}, braneworld scenarios~\cite{ParsaRiazi}, higher-dimensional gravity theories~\cite{higherworm,dynamic01}, hybrid metric-Palatini gravity~\cite{hybPala}, the pole dark energy model~\cite{poledark}, $f(R)$ gravity~\cite{FRDYNA} and other contexts.
\par 
In recent years, considerable  interest  has been focused on the subject of higher curvature gravity theories, much of which has been motivated through the attempts to provide a quantum description for gravitational interaction. Indeed, the quest for unifying the principles of quantum mechanics with gravitation has a long history and the first attempts to apply the standard quantization techniques to the Einstein-Hilbert action indicated that GR is nonrenormalizable~\cite{renormGR}. However, GR action becomes renormalizable when it is modified by higher curvature terms~\cite{ste1}. Therefore, if our starting point is a higher curvature classical theory of gravity instead of GR, we can get a renormalizable theory in which such higher curvature corrections can be regarded as candidates for quantum gravity. On the other side, from historical point of view, many efforts have been carried out in order to unify gravitational interaction with other fundamental ones. An approach to investigating this issue is to examine theoretical frameworks based on higher dimensions, i.e., beyond our conventional four-dimensional spacetime. In this respect, higher-dimensional gravity theories are taken into account as important ingredients of contemporary theories of fundamental physics, such as Kaluza-Klein, string theory, supergravity~\cite{kusts}, as well as holography~\cite{hol1} and cosmological scenarios~\cite{cos1}. Since the advent and development of these ideas, a great deal of effort has been expended to seek out physically reliable alternatives to GR theory. Of particular interest is {Lovelock} gravity, which is a natural generalization of GR to higher dimensions~\cite{love}. This theory is indeed the most general higher curvature gravity that possesses second-order equations of motion. The Lagrangian of {Lovelock} gravity is defined by a sum of dimensionally extended Euler densities so that in four dimensions all of the higher curvature correction terms appear as total derivatives, and thus, the theory reduces to GR. However, in higher dimensions, the new correction terms do make nontrivial contributions to the gravity sector of the action; see e.g.~\cite{revlove}, for recent reviews. Fortunately, in the framework of modified gravity theories the use of exotic matter can be avoided, thus providing  opportunities for traversable wormholes, and, among these theories, {Lovelock} gravity is not an exception. In this manner, static traversable wormholes have been introduced in third-order {Lovelock} theory~\cite{gmfl}, where the presence of {Lovelock} terms helps the energy conditions to be satisfied near the wormhole throat. Static wormholes in vacuum in higher-dimensional {Lovelock} gravity have been reported in~\cite{got2} and dynamic wormhole solutions in this framework with compact extra dimensions were analyzed in~\cite{riazi1}. {The occurrence of spacetime singularities has been also reported in the literature. In~\cite{Kobayashi2005}, exact radiating spacetimes filled with a null fluid have been found and application of these generalized Vaidya solutions in braneworld scenarios has been studied. Moreover, the gravitational collapse of a null dust fluid in Lovelock  gravity has been studied in~\cite{LoveVaidya} where it is shown that a naked singularity is formed whose nature and strength depend on spacetime dimensions or the power of the mass function. Also, in~\cite{Maeda2006}, the formation of massive naked singularities and their properties in a spherically symmetric dust cloud collapse has been investigated within a class of Lemaitre-Tolman-Bondi (LTB) spacetimes.}
\par
Motivated by the above results, we here seek for expanding wormhole configurations in higher dimensions in {Lovelock} gravity and check whether such dynamic wormholes are able to fulfill the energy conditions in arbitrary but finite time intervals. It is shown that higher-dimensional evolving wormholes can be obtained, satisfying the NEC throughout the spacetime~\cite{kolor}. However, in four dimensions, the solutions satisfy the NEC for a specific time interval. This paper is then organized as follows: In Sec.~(\ref{Gb}) we review the field equations of third-order {Lovelock} gravity and provide some preliminaries on dynamic wormholes. Sec. (\ref{WHS}) is devoted to the study of cosmological wormhole models and energy conditions in third-order {Lovelock} gravity. Finally, our conclusions are drawn in Sec. (\ref{concluding}). 
\section{Action and Field equations}\label{Gb}
The action in the framework of third-order {Lovelock} gravity is given by
\begin{equation}
	I=\int d^{n+1}x\sqrt{-g}\left( \mathcal{L}_{1}+\alpha _{2}^{\prime }\mathcal{L}%
	_{2}+\alpha _{3}^{\prime }\mathcal{L}_{3}\right) +{\mathcal S}_m,\label{Act1}
\end{equation}
where $\alpha_{2}^{\prime}$ and $\alpha _{3}^{\prime}$ are the second-(Gauss-Bonnet) and third-order {Lovelock} coefficients.
${\bf \mathcal{L}_{1}={\mathcal R}}$ is the Einstein-Hilbert Lagrangian, the term $\mathcal{L}_{2}$ is the Gauss-Bonnet Lagrangian given by
\begin{equation}
	\mathcal{L}_{2}={\mathcal R}_{\mu \nu \gamma \delta
	}{\mathcal R}^{\mu \nu \gamma \delta }-4{\mathcal R}_{\mu \nu }{\mathcal R}^{\mu \nu }+{\mathcal R}^{2},
\end{equation}
the third-order {Lovelock} Lagrangian $\mathcal{L}_{3}$ is defined as
\begin{eqnarray}
	\mathcal{L}_{3} &=&2{\mathcal R}^{\mu \nu \sigma \kappa }{\mathcal R}_{\sigma \kappa \rho \tau }{\mathcal R}_{\phantom{\rho \tau }{\mu \nu }}^{\rho \tau }+8{\mathcal R}_{\phantom{\mu \nu}{\sigma\rho}}^{\mu \nu }{\mathcal R}_{\phantom {\sigma \kappa} {\nu \tau}}^{\sigma \kappa }{\mathcal R}_{\phantom{\rho \tau}{ \mu \kappa}}^{\rho \tau }
	+24{\mathcal R}^{\mu \nu \sigma \kappa }{\mathcal R}_{\sigma \kappa \nu \rho }{\mathcal R}_{\phantom{\rho}{\mu}}^{\rho }+3{\mathcal R}{\mathcal R}^{\mu \nu \sigma \kappa }{\mathcal R}_{\sigma \kappa\mu \nu }
	\notag \\
	&&+24{\mathcal R}^{\mu \nu \sigma \kappa }{\mathcal R}_{\sigma \mu }{\mathcal R}_{\kappa \nu }+16{\mathcal R}^{\mu \nu
	}{\mathcal R}_{\nu \sigma }{\mathcal R}_{\phantom{\sigma}{\mu}}^{\sigma }
	-12{\mathcal R}{\mathcal R}^{\mu \nu }{\mathcal R}_{\mu \nu }+{\mathcal R}^{3} \,,
\end{eqnarray}%
and ${\mathcal S}_m$ contains the contribution due to matter fields. Now, varying action (\ref{Act1}) with respect to the metric, one gets the field equations as
\begin{eqnarray}
	G_{\mu \nu }^{\rm E}+\alpha _{2}^{\prime } G_{\mu \nu }^{\rm GB}+\alpha_{3}^{\prime }G_{\mu \nu }^{\rm TO}=\kappa_nT_{\mu \nu } \,, \label{Geq}
\end{eqnarray}
{where $\kappa_n=8\pi G_n$ with $G_n$ being the $n$-dimensional gravitational constant,} $T_{\mu \nu}$ is the {energy-momentum tensor} of matter field, $G_{\mu \nu}^{\rm E}$ is the Einstein tensor, and $G_{\mu \nu}^{\rm GB}$ and $G_{\mu \nu}^{\rm TO}$ are given by
\begin{eqnarray}
	G_{\mu \nu }^{\rm GB} &=&2({\mathcal R}{\mathcal R}_{\mu \nu }-{\mathcal R}_{\mu \sigma \kappa \tau
	}{\mathcal R}_{\phantom{\kappa \tau \sigma}{\nu}}^{\kappa \tau \sigma
}-2{\mathcal R}_{\mu \rho \nu \sigma }{\mathcal R}^{\rho \sigma
}-2{\mathcal R}_{\mu \sigma }{\mathcal R}_{\phantom{\sigma}\nu }^{\sigma })-\frac{1}{%
2}\mathcal{L}_{2}g_{\mu \nu } \\
G_{\mu \nu }^{\rm TO} &=&-3(4{\mathcal R}^{\tau \rho \sigma \kappa }{\mathcal R}_{\sigma
	\kappa\lambda\rho}{\mathcal R}_{\phantom{\lambda }{\nu \tau \mu}}^{\lambda }-8{\mathcal R}_{%
	\phantom{\tau \rho}{\lambda \sigma}}^{\tau \rho}{\mathcal R}_{\phantom{\sigma
	\kappa}{\tau \mu}}^{\sigma \kappa }{\mathcal R}_{\phantom{\lambda }{\nu \rho \kappa}%
}^{\lambda }+2{\mathcal R}_{\nu }^{\phantom{\nu}{\tau \sigma\kappa}}{\mathcal R}_{\sigma \kappa
\lambda \rho}{\mathcal R}_{\phantom{\lambda \rho}{\tau \mu}}^{\lambda \rho } \nonumber\\
&&-{\mathcal R}^{\tau \rho \sigma \kappa }{\mathcal R}_{\sigma \kappa \tau \rho }{\mathcal R}_{\nu \mu }+8{\mathcal R}_{%
	\phantom{\tau}{\nu \sigma \rho}}^{\tau }{\mathcal R}_{\phantom{\sigma \kappa}{\tau \mu}%
}^{\sigma \kappa}{\mathcal R}_{\phantom{\rho}\kappa }^{\rho }+8{\mathcal R}_{\phantom{\sigma}{\nu \tau \kappa}}^{\sigma }{\mathcal R}_{\phantom {\tau \rho}{\sigma \mu}%
}^{\tau \rho }{\mathcal R}_{\phantom{\kappa}{\rho}}^{\kappa } \nonumber\\
&&+4{\mathcal R}_{\nu }^{\phantom{\nu}{\tau \sigma \kappa}}{\mathcal R}_{\sigma \kappa
	\mu \rho }{\mathcal R}_{\phantom{\rho}{\tau}}^{\rho }-4{\mathcal R}_{\nu
}^{\phantom{\nu}{\tau \sigma \kappa }}{\mathcal R}_{\sigma \kappa \tau \rho
}{\mathcal R}_{\phantom{\rho}{\mu}}^{\rho }+4{\mathcal R}^{\tau \rho \sigma \kappa
}{\mathcal R}_{\sigma \kappa \tau \mu }{\mathcal R}_{\nu \rho
}+2{\mathcal R}{\mathcal R}_{\nu }^{\phantom{\nu}{\kappa \tau \rho}}{\mathcal R}_{\tau \rho \kappa \mu } \nonumber\\
&&+8{\mathcal R}_{\phantom{\tau}{\nu \mu \rho }}^{\tau }{\mathcal R}_{\phantom{\rho}{\sigma}%
}^{\rho}{\mathcal R}_{\phantom{\sigma}{\tau}}^{\sigma}-8{\mathcal R}_{\phantom{\sigma}{\nu \tau
\rho }}^{\sigma }{\mathcal R}_{\phantom{\tau}{\sigma}}^{\tau }{\mathcal R}_{\mu }^{\rho }-8{\mathcal R}_{%
\phantom{\tau }{\sigma \mu}}^{\tau \rho }{\mathcal R}_{\phantom{\sigma}{\tau
}}^{\sigma}{\mathcal R}_{\nu \rho }-4{\mathcal R}{\mathcal R}_{\phantom{\tau}{\nu \mu \rho }}^{\tau }{\mathcal R}_{\phantom{\rho}%
\tau }^{\rho } \nonumber\\
&&+4{\mathcal R}^{\tau \rho }{\mathcal R}_{\rho \tau }{\mathcal R}_{\nu \mu
}-8{\mathcal R}_{\phantom{\tau}{\nu}}^{\tau}{\mathcal R}_{\tau \rho }{\mathcal R}_{\phantom{\rho}{\mu}}^{\rho }+4{\mathcal R}{\mathcal R}_{\nu \rho }{\mathcal R}_{%
\phantom{\rho}{\mu }}^{\rho }-{\mathcal R}^{2}{\mathcal R}_{\nu \mu })-\frac{1}{2}\mathcal{L}%
_{3}g_{\mu \nu }.
\end{eqnarray}
In Lovelock theory, for each Euler density of order $\bar{k}$ in $n$-dimensional spacetime, only the terms with $\bar{k}<n$ {contribute to} the equations of motion~\cite{dehgan1}. {Therefore, the allowed solutions of the Lovelock
theory are derived in $n\geq 7$ dimensions}. Note that action (\ref{Act1}) is {recovered} in the low energy limit of string theory \cite{myg}.

In this work, we consider $n$-dimensional traversable wormhole spacetimes, by replacing the two-sphere~\cite{mt} in the angular part of the metric with an $(n-2)$-sphere. This gives the following form for the line element
\begin{equation}
	ds^{2}=-e^{2\phi (r)}dt^{2}+ R\left(t\right)^2\left[\frac{dr^{2}}{1-b(r)/r} +r^{2}d\Omega
	_{n-2}^{2}\right]\,,
	\label{evw}
\end{equation}
where $d\Omega_{n-2}^{2}$ is the metric on the surface of the $(n-2)$-sphere, $R(t)$ is the scale factor of the universe, $\phi(r)$ is the redshift function as it is related to the gravitational redshift, and $b(r)$ is the wormhole shape function. The shape function must satisfy the flare-out condition at the throat~\cite{mt}, i.e., we must have $b^{\prime}(r_0)<1$ and $b(r)<r$ for $r>r_0$ in the whole spacetime, where $r_0$ is the throat radius. The condition $\phi(r)=0$ has been discussed in~\cite{mouders} that zero-tidal force wormholes are supported by anisotropic fluid with a diagonal energy momentum tensor. Our aim here is to study evolving wormholes with anisotropic pressures in an inhomogeneous spacetime which merge smoothly to the homogeneous FLRW model. In the present work, we consider $\phi(r)=0$ in order to ensure the absence of horizons and singularities throughout the spacetime. These evolving Lorentzian wormholes are conformally related to another family of static wormholes with zero-tidal force. The general constraints on these functions have been discussed by Morris and Thorne in~\cite{mt}. It is clear that if $b(r)$ and $\phi(r)$ tend to zero the metric~(\ref{evw}) reduces to a flat {FLRW} metric, and as $R(t) \rightarrow constant$ the static Morris-Thorne wormhole is recovered. In the model herein, we seek a way to determine the shape function $b(r)$ and the scale factor $R(t)$ in order to construct dynamical wormhole configurations. {We use a unit system with $\kappa_n=1$}.
\par

In an orthonormal reference frame, the nonzero components of the {energy-momentum tensor} read
\begin{equation}
{T}^{\,\,\mu}_{\nu}={\rm
	diag}\left[-\rho\left(r,t\right),P_{r}\left(r,t\right),P_{t}\left(r,t\right),P_{t}\left(r,t\right)
,...\right],
\end{equation}
where $\rho\left(r,t\right)$ is the energy density and $P_{r}\left(r,t\right)$ and $P_{t}\left(r,t\right)$ are the radial and transverse pressures, respectively. Thus, the gravitational field equation (\ref{Geq}) provides us with the components of ${T}^{\,\,\mu}_{\nu}$ as
	\begin{eqnarray}
	\rho(r,t)&=&{\frac { \left[  \left( n-1 \right)R ^{2}{H}^{2}+ \left( n-3\right)Q+2p\right]\left(n-2\right)}{2 R^{2}}}
	+\notag \\&&{\frac {\alpha\left(Q+H^{2} R^{2}\right)\left( n-2 \right)\left[\left( n-1 \right)R^{2}H^{2}+ \left( n-5 \right) Q +4p\right] }{2R^{4}}}+\notag \\&& {\frac {\beta\left( {R}^{2}{H}^{2}+Q \right) ^{2} \left( n-2 \right)\left[{H}^{2}\left(n-1 \right)R^{2}+ \left( n-7 \right) Q+6p\right]}{2R^{6}}},\label{feild1}
	\end{eqnarray}

\begin{eqnarray}
P_{r}\left(r,t\right)&=&-{\frac { \left[ 2\, \dot{H} R ^{2}+ \left( n-1 \right) 
		 R ^{2}  H ^{2}+ \left( n-3 \right) Q \right]  \left( 
		n-2 \right) }{2 R ^{2}}}-\notag \\
&&{\frac{\left[4\dot{H}R^{2}+ \left( n-1 \right)R^{2} H^{2}+\left(n-5\right)Q\right] 
		\left( n-2 \right) \alpha\, \left(Q+H^{2}R^{2} \right)}{ 2R^{4}}}\notag \\
&&-{\frac{\left[6{R}^{2}\dot{H}+R^{2} \left( n-1 \right)H^{2}+\left( n-7 \right) Q \right]\left(Q+ H^{2}{R}^{2} \right)^{2}\beta\left( n-2 \right) }{{2R}^{6}}},
 \label{feild2}
\end{eqnarray}
\begin{eqnarray}
P_{t}\left(r,t\right)&=&{\frac {-2R ^{2} \left(n-2\right)\dot{H} - H^{2} \left( n-1 \right) \left( n-2 \right) 
		R ^{2}- \left(  \left( n-4 \right) Q +2p\right)  \left( n-3 \right) }{2R^{2}}}+\notag \\
&&\frac{\alpha \left[-4{R}^{2} \left(  \left( n-2 \right) {H^{2}}{R}^{2}+ \left( n-4 \right) Q+2p \right)\dot{H} -H ^{4} \left(n-1 \right)  \left( n-2 \right) {R}^{4}\right]}{2R^{4}}+\notag \\&&\frac{\alpha(-2\left(  \left( n-4 \right) Q+2p \right) {H}^{2} \left( n-3
	\right) {R}^{2}- \left(  \left( n-6 \right) Q+4p \right) Q \left(n-5 \right))}{2R^{4}}-\notag \\&&{\frac {\beta\left[ 2{R}^{2} \left( n-2 \right)\dot{H} +{R}^{2} \left( n-1 \right) \left(n-2\right)H^{2}+ 3\left( n-3 \right) 
		\left( Qn-4Q+2p \right)\right] H^{4}}{{2R}^{2}}}+\notag \\&& {\frac {\beta\left[-12\, \left(  \left( n-4 \right) Q+2p \right) {R}^{4}\dot{H}-3Q \left(n-5\right)\left(\left( n-6 \right) Q+4p \right){R}^{2}\right]H^{2}}{{2R}^{6}}}+\notag \\&&{\frac {\beta\left[-6Q \left(\left(n-6\right)Q+4p \right)R^{2}\dot{H}-{Q}^{2}\left(n-7\right)\left(\left( n-8 \right)Q+6p \right)\right]}{{2R}^{6}}}
,
\label{feild3}
 \end{eqnarray}
where an overdot denotes a derivative with respect to time. We define  $\alpha=(n-3)(n-4)\alpha^\prime_2$, $\beta=(n-3)(n-4)(n-5)(n-6)\alpha^\prime_3$   and $H=\frac{\dot{R}(t)}{R(t)}$ for notational convenience and the functions $p$ and $Q$ are given by
\begin{eqnarray}
p={\frac {b^{\prime} r-b}{2{r}^{3}}},~~~~Q=\frac{b}{r^3}.
\end{eqnarray}
One can check that for $R(t)=constant$ and $\beta=0$ Eqs. (\ref{feild1})-(\ref{feild3}) reduce to the field equations as derived in the paper by Bhawal and Kar~\cite{bhakar}. It is also easy to check that, for $\alpha=0$ and $\beta=0$ the field equations become those of higher-dimensional evolving wormholes in Einstein gravity~\cite{kolor}.

\section{WORMHOLE SOLUTIONS}\label{WHS}
\subsection{ENERGY CONDITIONS}
It is well-known that static traversable wormholes in four dimensions violate energy conditions~\cite{hoc1} which is due to the fulfillment of the flaring-out condition near the throat of the wormhole. However, the energy conditions can be satisfied in the
vicinity of static wormhole throats in the framework of higher-dimensional gravity theories~\cite{dehha2} and the whole spacetime in the
case of higher-order curvature terms~\cite{tari}. On the other hand, evolving wormholes may provide a setting to avoid the violation of energy conditions for a limited time period. For the sake of physical reasonability of wormhole configuration the WEC must be satisfied. This condition requires that $T_{\mu \nu }U^{\mu }U^{\nu}\geq 0$, where $U^{\mu }$ is a timelike vector field. For a diagonal {energy-momentum tensor}, the WEC leads to the following inequalities
\begin{equation}\label{E11}
\rho \geq0      , \ \ \ \ \ \rho+P_{r} \geq0,\ \ \ \ \ \and \rho+P_{t}\geq0.
\end{equation} 
Note that the last two inequalities are defined as the NEC. Using Eqs.~(\ref{feild1})-(\ref{feild3}), one finds the following relationships:
\begin{eqnarray}
\rho +P_{r}&=&{\frac { \left(p- \dot{H} {R}
^{2} \right)  \left( n-2 \right) }{R^{2}}}-{\frac{ 2\alpha\left(  \dot{H}R^{2}-p \right)\left(n-2 \right)\left(Q+H ^{2}R^{2} \right) }{R^{4}}}-\notag \\&&{\frac {3\beta \left( Q+ H ^{2}R^{2}\right) ^{2} \left(R^{2}\dot H -p\right)  \left( n-2 \right)}{R^{6}}}
,
\label{EGBNEC}
\end{eqnarray}%

\begin{eqnarray}
 	\rho +P_{t}&=&{\frac {-R^{2} \left( n-2 \right) \dot{H} 
 			+ \left( n-3 \right) Q+p}{{R}^{2}}}+\frac{2\alpha\left(  \left(n-5 \right) Q+3\,p \right) Q}{R^4}+\notag \\
 	&&{\frac {2\alpha\left[ -{R}^{2} \left(H ^{2} \left( n-2 \right) {R}^{2}+ \left( n-4 \right) Q+2p
 			\right)\dot{H} + \left(  \left( n-3\right) Q+p \right) H^{2}R^{2} \right] }{R^{4}}}-\notag\\&&{\frac {\beta\left[3H ^{4} \left( n-
 			  2 \right) R^{4}+R^{2} \left(  6\left( n-4 \right) Q+p \right) H ^{2}+Q \left( \left( n-6 \right) Q+4p \right)\right]\dot{H} }{R^{4}}}+\notag\\&&{\frac{3\beta\left( Q+H^{2}R^{2}  \right)  \left[ {R}^{2} \left(  \left( n-3 \right) Q+p \right) H^{2}+ \left(  \left( n-7 \right) Q +5p \right) Q \right] }{R^{6}}},
 		 	\end{eqnarray}
where a prime and an overdot stand for differentiation with respect to $r$ and $t$, respectively. From Eq. (\ref{EGBNEC}) we get at the throat
\begin{eqnarray}
 \left( \rho + P_r \right)\Big|_{r=r_0}=- \left( n-2
 \right)\left({\frac {1-b^{\prime}_0}{2R^{2}r_0^{2}}}+\dot{H}\right) \left(1+2\alpha\,{H}^{2}+3\beta{H}^4+{\frac {2\alpha}{r_0^{2}R^{2}}}+{\frac{3\beta}{r_0^{4}R^{4}}}+{\frac {6\beta H^2}{r_0^{2}R^{2}}}\right),\label{rhoprr0}
\end{eqnarray}
 which shows that for $\alpha=0$, $\beta=0$, and $H=constant$ the NEC, and consequently the WEC, are violated at the throat, due to the flaring-out condition. In order to satisfy $\rho + P_r >0$ in Lovelock gravity, one can choose suitable values of $\alpha$, $\beta$, and $H$ at the wormhole throat. 
\subsection{COSMOLOGICAL WORMHOLES}
We now have three equations, namely the field equations $(\ref{feild1})-(\ref{feild3})$, with the five unknown functions: $\rho(r,t),P_r(r,t),P_t(r,t),b(r)$, and $R(t)$. Therefore, in order to
determine the wormhole geometry, one can adopt several strategies~\cite{stra1}. Here, we are interested in studying evolving
wormholes with anisotropic pressures in an inhomogeneous spacetime, which merge smoothly to the cosmological background. The wormhole solutions presented in a cosmological background  have the interesting property that their Ricci scalar is independent
of the radial coordinate $r$, similar to what happens in the cosmological setting~\cite{ebra}. In other words, the scalar curvature of the
spacetime is a function of time, only. The Ricci scalar corresponding to the metric~(\ref{evw}) will play a fundamental role in our analysis, which is obtained as
 \begin{align}
\mathcal{R}(t,r) = \left(n-1\right)\left(nH^2+2\dot{H}\right)+{\frac{\left( n-2 \right)\left[\left( n-3 \right) Q+2p\right]}{R(t)^{2}}}.
\end{align}
It can be seen that the second term depends on the $r$ coordinate, and hence in a cosmological background condition ${\frac {\partial }{\partial r}}{\it \mathcal{R}} \left( t,r \right)=0 $ leads to the following differential equation:
\begin{align}
\left( n-3 \right) {\frac {d}{dr}}Q \left( r \right) +2{\frac {d}{dr}}p \left( r \right) =0.\label{difer1}
\end{align}
The above differential equation  provides us with the following form for the shape function:
\begin{align}
b \left( r \right) ={C_1}r^{3}+{C_2}r^{4-n},
\end{align}
where $ C_1$ and $ C_2$ are constants of integration. We note that the space slice $t = constant$ of the metric~(\ref{evw}) for the shape function introduced (with $C_1=0$) coincides with the space slice of the $n$-dimensional extension of	the Schwarzschild black hole~\cite{maper}. Using the condition $b(r_0)=r_0$ at the throat, we get
\begin{align}
b(r)={C_1}r^{3}-r_0^{n-3} \left(C_1r_0^{2}-1 \right)r^{4-n}. \label{shape1}
\end{align}
Also the condition $b^{\prime}(r_0) < 1$ leads to the following inequality:
\begin{align}
{C_1}<{\frac {n-3}{r_0^{2} \left( n-1 \right) }}.
\end{align}
We can now obtain the constant $C_1$ using the fact that the spacetime is asymptotically FLRW along with applying the normalization $C_1= 0, \pm 1$ for the curvature constant. It is clear that solutions with $C_1= 0$ (flat universe) are asymptotically flat, i.e., $\frac{b(r)}{r}$ tends to zero as $r\rightarrow\infty$. Also, the condition $b^{\prime}(r_0) < 1$ is satisfied for the solution with $C_1= -1$ (open universe). For the case of the wormhole solution with $C_1= 1$ (closed universe), the wormhole configuration cannot be arbitrarily large. {It is worth mentioning that other curvature invariants, such as the Kretschmann scalar, ${\sf K}={\mathcal R}^{\alpha\beta\gamma\delta}{\mathcal R}_{\alpha\beta\gamma\delta}$, and the Weyl square, ${\sf C}^2=C^{\alpha\beta\gamma\delta}C_{\alpha\beta\gamma\delta}$ or combinations of them, can be utilized in order to obtain nontrivial wormhole solutions. The Kretschmann scalar for the spacetime metric (\ref{evw}) is given by
	\be\label{KRETC}
	{\sf K}=\f{2(n-2)}{R^4}\left[(n-3)Q^2+2p^2+(4p+(2n-6)Q)\dot{R}^2\right]+\f{2(n-1)}{R^4}\left[(n-2)\dot{R}^4+2R^2\ddot{R}^2\right].
	\ee 
	We then observe that the Kretschmann invariant cannot be separated into time and radial dependent functions, hence, finding wormhole solutions using this invariant may not be as simple as utilizing the Ricci scalar. For the case of the Weyl square we get
	\be
	{\sf C}^2=\f{4(n-3)\left(p-Q\right)^2}{(n-1)R^4},
	\ee
	where we see that since ${\sf C}^2$ is separable in $r$ and $t$ one can find nontrivial solutions for the shape function, assuming a suitable form for the $(t,r)$-dependence of the Weyl square invariant.}
 \par
With $b(r)$ given by Eq. (\ref{shape1}) along with using the field equations (\ref{feild1})-(\ref{feild3}), we obtain 
 	\begin{eqnarray}
 	\rho(r,t)&=&\rho_{cb}(t)
 	-\notag \\&&{\frac { \left( n-1 \right)  \left( n-2 \right) \alpha r_0^{-4} \left(C_1r_0^{2}-1\right)^{2}
 			}{2R^{4}}}\left(\frac{r_0}{r}\right)^{2n-2}-\notag\\&&\,{\frac {3\beta\left( n-1 \right)\left({C_1}{ r_0}^{2}-1\right)^{2}\left( n-2 \right) {r_0}^{-4}\left(R^{2}{H}^{2}+{C_1} \right) }{2R^{6}}}\left(\frac{r_0}{r}\right)^{2n-2}
 	+\notag\\&& {\frac {\beta\left( n-1 \right)\left({C_1}{r_0}^{2}-1\right)^{3}\left( n-2 \right)r_0^{-6}}{R^{6}}} \left(\frac{r_0}{r}\right)^{3n-3},\label{rho1f}
 	\end{eqnarray}
 	
 	\begin{eqnarray}
 	P_{r}\left(r,t\right)&=&P_{cb}(t)
 	+\notag \\
 	&&{\frac { \left( n-2 \right) r_0^{-2} \left[2\alpha R ^{2}\dot{H} +\left(\alpha{H}^{2}+\frac{1}{2} \right)  \left( n-3 \right) {R}^{2}+\alpha{C_1}\left(n-5\right)  \right] \left( {C_1}r_0^{2}-1\right) }{R^{4}}}\left(\frac{r_0}{r}\right)^{\,n-1}-\notag \\&&{\frac {\alpha\,\left( {C_1}{{r_0}}^{2}-1 \right)^{2} \left(n-5 \right)\left( n-2 \right)}{2R^{4}}}\left(\frac{r_0}{r}\right)^{2n-2}+\notag \\&&{\frac {3\beta \left[\left(  \left( n-3 \right) {H}^{2}+4\,{\dot H} \right) {R}^{2}+ \left( n-7 \right) {C_1} \right]
 			\left( n-2 \right) r_0^{-2} \left( {R}^{2}{H}^{2}+{C_1}\right)  \left( {C_1}r_0^{2}-1 \right) }{2{R}^{6}}}\left(\frac{r_0}{r}\right)^{n-1}-\notag\\&&{\frac {3\beta \left( n-2 \right) r_0^{-4} \left( {C_1}r_0^{2}-1 \right) ^{2} \left[\left(n-7\right) {C_1}+{R}^{2} \left(2{\dot H}+{H}^{2}n-5{H}^{2}
 			\right)\right] }{2{R}^{6}}}\left(\frac{r_0}{r}\right)^{2n-2}+\notag\\&&{\frac{r_0^{-6} \left({C_1}r_0^{2}-1\right)^{3} \left( n-7 \right)\left( n-2 \right) \beta}{2R^{6}}}\left(\frac{r_0}{r}\right)^{3n-3},
 \label{prr2}
 	\end{eqnarray}
 	\begin{eqnarray}
 	P_{t}\left(r,t\right)&=&P_{cb}(t)-
 	\notag \\
 	&&{\frac{\left[2\alpha R ^{2}\dot{H} 
 			+\left( \alpha{H}^{2}+\frac{1}{2} \right)\left(n-3\right) {R}^{2}+\alpha{C_1}\left( n-5\right)  \right] 
 			\left( {C_1}r_0^{2}-1 \right)}{R^{4}}}\left(\frac{r_0}{r}\right)^{n-1}+\notag \\ &&{\frac {n\alpha\left( {C_1}r_0^{2}-1 \right) ^{2} \left( n-5 \right)}{2{R}^{4}}}\left(\frac{r_0}{r}\right)^{2n-2}-\notag\\&&{\frac{3\beta \left[  \left(  \left( n-3 \right) {H}^{2}+4\,{\dot H} \right) {R}^{2}+ \left( n-7 \right) {C_1} \right]\left( {C_1}r_0^{2}-1 \right)  \left(R^{2}H^{2}+{C_1} \right)r_0^{-2}}{2{R}^{6}}}\left(\frac{r_0}{r}\right)^{n-1}-\notag\\&&
 	{\frac {3\beta \left( {C_1}r_0^{2}-1\right) ^{2}nr_0^{-4} \left[\left( 2{\dot H}+n{H}^{2}-5{H}^{2} \right) {R}^{2}+ \left( n-7 \right) {C_1} \right] }{2{R}^{6}}}\left(\frac{r_0}{r}\right)^{2n-2}-\notag\\&&{\frac { \left( {C_1}r_0^{2}-1 \right)^{3}{r_0}^{-6}\beta\left( n-7 \right)  \left( n-1/2 \right) }{R^{6}}}\left(\frac{r_0}{r}\right)^{3n-3},\label{ptt3}
 		\end{eqnarray}
 		where the $\rho_{cb}$ and $P_{cb}$ components correspond with the cosmological background and are given by
 		\begin{align}
 		\rho_{cb}(t)={\frac { \left( n-2 \right) \left( n-1 \right) \left( {H}^{2}{R}^{2}+{C_1}
 				\right) \left[ \left(H^{4}R^{4}+2{C_1}H^{2}R^{2}+C_1^{2} \right) \beta+ \left(H^{2}R^{4}+{C_1}R^{2} \right) \alpha+R^{4}\right]}{2R^{6}}},
 		\end{align}
 		\begin{eqnarray}
 	P_{cb}(t)&=&-{\frac { \left( \left[\left( 2+4\alpha H ^{2} \right) R^{4}+4\alpha R^{2}{C_1}
 			\right]\dot{H} + H ^{2} \left( 1+\alpha H ^{2} \right)  \left( n-1 \right)R^{4} \right)\left( n-2 \right) }{2R^{4}}}\notag \\
 	&&-\frac{\left[\left(n-3\right) {C_1}\left( 1+2\alpha H ^{2}\right)R^{2}+\alpha{C_1}^{2} \left(n-5
 		\right)\right]\left(n-2\right) }{2{R}^{4}}-\notag\\&& {\frac {\beta\, \left( n-2 \right)  \left[ 6\,{R}^{2}{\dot H}+{
 				H}^{2} \left( n-1 \right) {R}^{2}+\left( n-7 \right) {C_1}\right]\left(R^{2}H^{2}+{C_1} \right)^{2}}{2{R}^{6}}}
 	.
 \end{eqnarray}
Notice that for our solutions in a cosmological background, the components of $\rho$, $P_r$, and $P_t$ are asymptotically independent of $r$. Moreover, their first terms depend only on time corresponding to a cosmological background as described by FRW spacetime. Let us now investigate the features of the evolving wormhole. We can determine the behavior of the scale factor by applying a linear equation of state between the radial pressure and energy density of the cosmological background profiles, i.e., $P_{cb}=w\rho_{cb}$. We then obtain
 \begin{eqnarray}
 & R^{2} \left(2\dot{H}R^{2}+ H ^{2} \left( w+1 \right)\left( n-1 \right) R ^{2}+ \left(\left( w+1 \right) n-w-3 \right) {C_1} \right) +\notag \\ &\alpha\left( {C_1}+  H^{2} R^{2} \right)  \left[ 4w {\dot{H}}R ^{2}+ H^{2} \left( w+1 \right) \left( n-1 \right) R^{2}+ \left(\left( w+1 \right) n-w-5 \right) {C_1} \right]+\notag\\& \left[ 6w{\dot H}R^{2}+H^{2} \left( 1+w \right)  \left( n-1 \right)R^{2}+ \left(  \left( n-7 \right) w-1+n \right) {C_1} \right] \left( {C_1}+{R}^{2}{H}^{2} \right) ^{2}\beta =0.
 \label{dynamic1}
 \end{eqnarray}
One can check that for $\beta=0$, the solution of Eq.~(\ref{dynamic1}) reduces to the scale factor for a higher dimensional in Gauss-Bonnet gravity~\cite{dymeh20}. In the following subsections, with the help of the master equation~(\ref{dynamic1}), we will
 determine the behavior of the scale factor and the related properties of the energy conditions for the wormhole
 geometry in the presence of Lovelock gravity. Thus, in order to study an evolving wormhole in
 detail, we consider three cases: $C_1=0$ and $C_1=\pm 1$.
 
\subsection{Solutions for the case $C_1=0$}
\begin{figure}
	\begin{center}
		\includegraphics[scale=0.42]{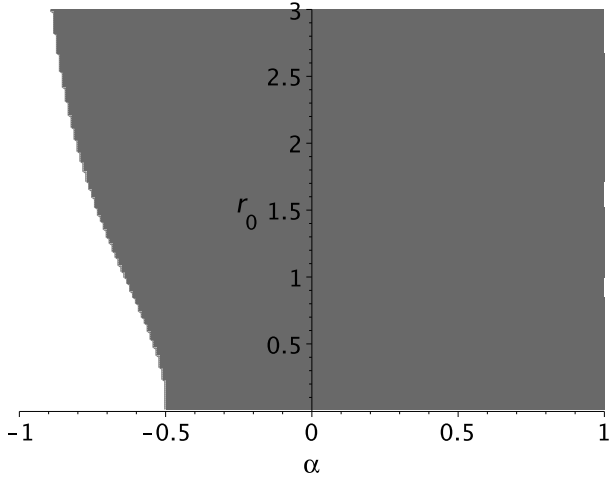}
		\caption {The allowed region for the $\alpha$ parameter and wormhole throat for $R_2=1$, $h=1$, and $n=7$.}\label{fggc0a}
	\end{center}
\end{figure}

\begin{figure}
	\begin{center}
		\hspace{-0.7cm}
		\includegraphics[scale=0.4]{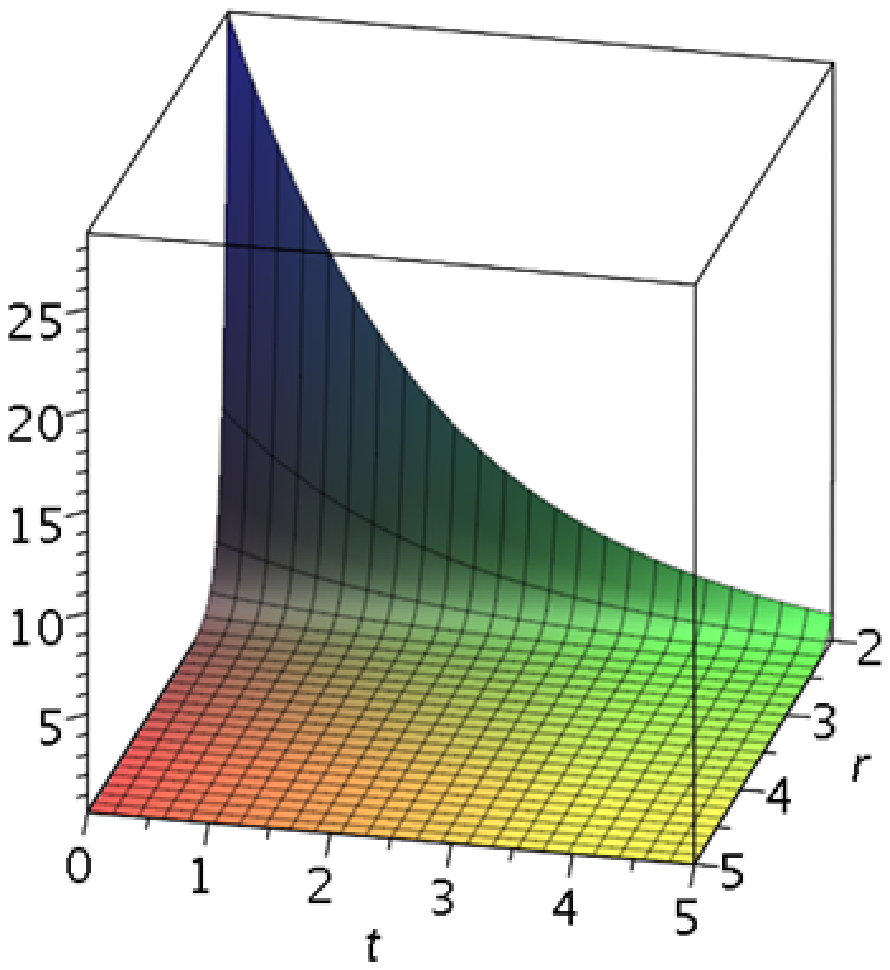}\hspace{-1.5cm}
		\includegraphics[scale=0.4]{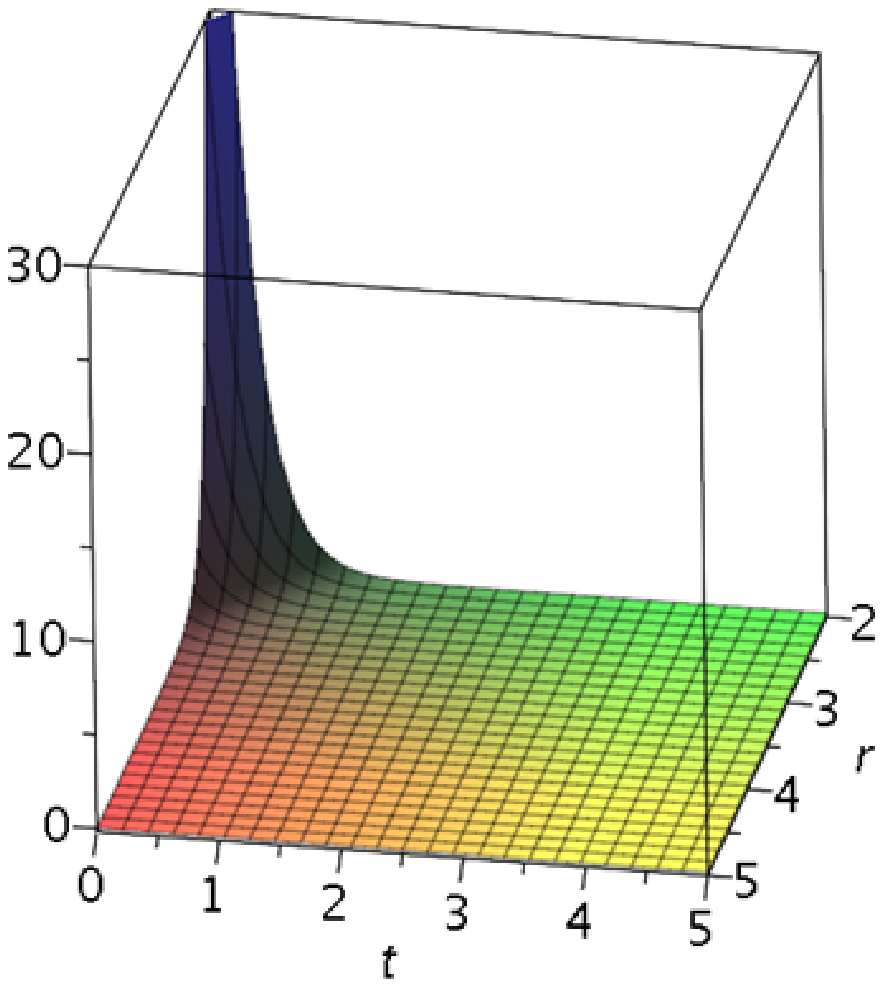}\hspace{-1.5cm}
		\includegraphics[scale=0.4]{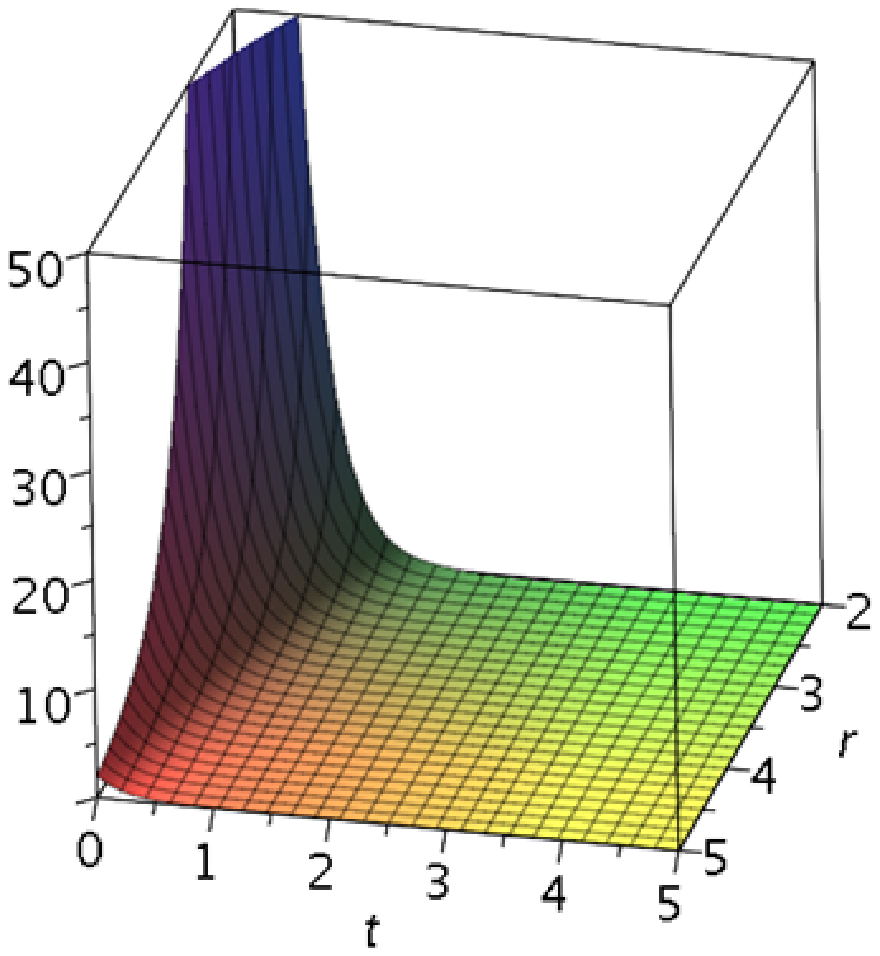}\hspace{-0.4cm}\vspace{-1.8cm}
		\caption {The behavior of $\rho$, $\rho +p_{r}$  and $\rho +p_{t}$ versus $r$ and $t$, respectively, from left to right, for $w=-1$, $r_0=2$, $\alpha=1$, $\beta=-1$, ${R_2=0.5}$, and ${h=1.5}$ in  seven dimensions.}\label{fggc0}
	\end{center}
\end{figure}

We first try to solve the differential equation~(\ref{dynamic1}) for  the inflationary expanding regime, i.e.,  $w=-1$. We then obtain the scale factor as $R(t)=R_{2} {\rm e}^{ht}$, where $R_2$ and $h$ are real constants. In order to check the WEC we rewrite expressions~(\ref{E11}) for this solution, as
\begin{eqnarray}
 \rho(r,t)&=& \frac{\left( n-2 \right) \left( n-1 \right) {h}^{2} \left( 1+\alpha{h}^{2}+\beta h^4 \right)}{2}-{\frac{(\alpha+3\beta h^2)\left( n-1 \right) \left( n-2 \right)}{2R_2^{4}r_0^{4}{{\rm e}^{4ht}}}}\left(\frac{r_0}{r}\right)^{2n-2}-\notag\\&&{\frac{\left( n-2 \right) \left( n-1 \right)\beta}{r_0^{6}R_2^{6}{\rm e}^{6ht}} }\left(\frac{r_0}{r}\right)^{3n-3},
 \end{eqnarray}
 \begin{eqnarray}
 \rho+P_r&=&-{\frac { \left( n-2 \right) \left( n-3 \right) \left( 1+2\alpha{h}^{2}+3\beta h^4 \right)}{2R_2^{2}r_0^{2}{{\rm e}^{2ht}}
 	}}\left(\frac{r_0}{r}\right)^{n-1}-{\frac { \left( n-2 \right)\left( n-3 \right) \left(\alpha+3\beta h^2\right)
 	}{R_2^{4}r_0^{4}{{\rm e}^{4ht}}}}\left(\frac{r_0}{r}\right)^{2n-2}-\notag\\&&{\frac { 3\beta\left( n-2 \right)  \left( n-3 \right)}{2r_0^{6}R_2^{6}{{\rm e}^{6ht}}}}\left(\frac{r_0}{r}\right)^{3n-3}
,
\end{eqnarray}

 \begin{eqnarray}
\rho+P_t&=&{\frac {\left( n-3 \right)  \left( 1+2\alpha{h}^{2}
		+3\beta h^4\right) }{2R_2^{2}r_0^2{{\rm e}^{2ht}}}}\left(\frac{r_0}{r}\right)^{n-1}-{\frac {\left(\alpha+3\beta h^2\right)\left( 1+n \right) }{R_2^{4}r_0^{4}{{\rm e}^{4ht}}}}\left(\frac{r_0}{r}\right)^{2n-2}-\notag\\&&{\frac {3\beta \left(3n-1\right)}{2r_0^{6}R_2^{6}{{\rm e}^{6ht}}}}\left(\frac{r_0}{r}\right)^{3n-3}
.
 \end{eqnarray}
It is clear that both $\rho+P_r$ and $\rho+P_t$ tend to zero as $t\to\infty$, with opposite signs. Therefore, in the limit of large times, one of the $\rho+P_r$ or $\rho+P_t$ quantities is negative and consequently the WEC is violated. However, {it is seen} that one can {set the coefficient of the ${\mathcal L}_3$ term as,} $\beta={-\frac {1+2\alpha{h}^{2}}{3{h}^{4}}}$, such that the first term is eliminated. For this case, the values of these quantities at the throat for $t=0$ are given by
\begin{align}
\rho(r)\Big|_{r=r_0}={\frac { \left( n-2 \right)  \left( n-1 \right)  \left( 2+{h}^{8}
		\alpha R_2^{6}r_0^{6}+2{h}^{6}R_2^{6}r_0^{6}+3\alpha{h}^{4}R_2^{2}r_0^{2}+\left( 3R_2^{2}r_0^{2}+4\alpha \right){h}^{2} \right) }{6{h}^{4}R_2^{6}r_0^{6}}},
\end{align}

\begin{align}
\rho(r)+P_r(r)\Big|_{r=r_0}={\frac { \left( 1+2\alpha{h}^{4}R_2^{2}{r_0}^{2}+2\left(R_2^{2}{{r_0}}^{2}+\alpha\right) {h}^{2} \right)\left( n-3 \right)\left( n-2 \right) }{{h}^{4}R_2^{6}r_0^{6}}},
\end{align}
and
\begin{align}
\rho(r)+P_t(r)\Big|_{r=r_0}={\frac {2\alpha R_2^{2}r_0^{2} \left(1+n\right) {h}^{4}+ \left( 2r_0^{2} \left( 1+n \right) R_2^{2}+ \left( 6n-2 \right) \alpha \right) {h}^{2}+3n-1}{2{h}^{4} R_2^{6}r_0^{6}}}.
\end{align}
%
Figure (\ref{fggc0a}) shows that it is possible to choose suitable values for the {$\alpha$ parameter}
in order to satisfy the WEC at the throat. Also, with increasing the value of $r_0$ the WEC {is} satisfied by choosing larger values of $\mid{\alpha}\mid$. {However, it is still possible to choose appropriate values of the $\alpha$ parameter and, as Fig.~(\ref{fggc0}) shows, the WEC is satisfied at all times and $r\geq r_0$}.
 \subsubsection{Numerical solutions  for the case $C_1=0$}
\begin{figure}
	\begin{center}
		\includegraphics[scale=0.35]{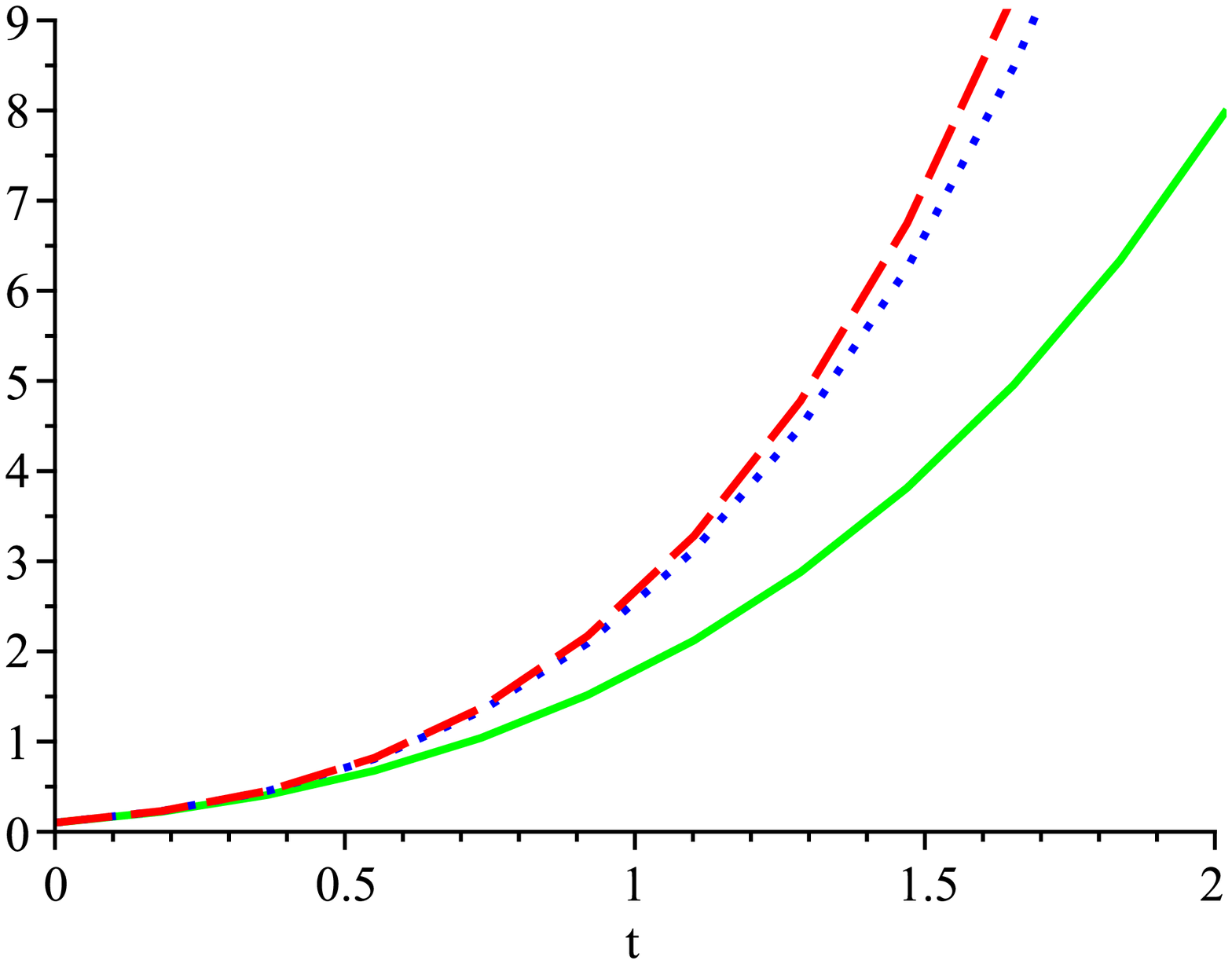}
		\includegraphics[scale=0.35]{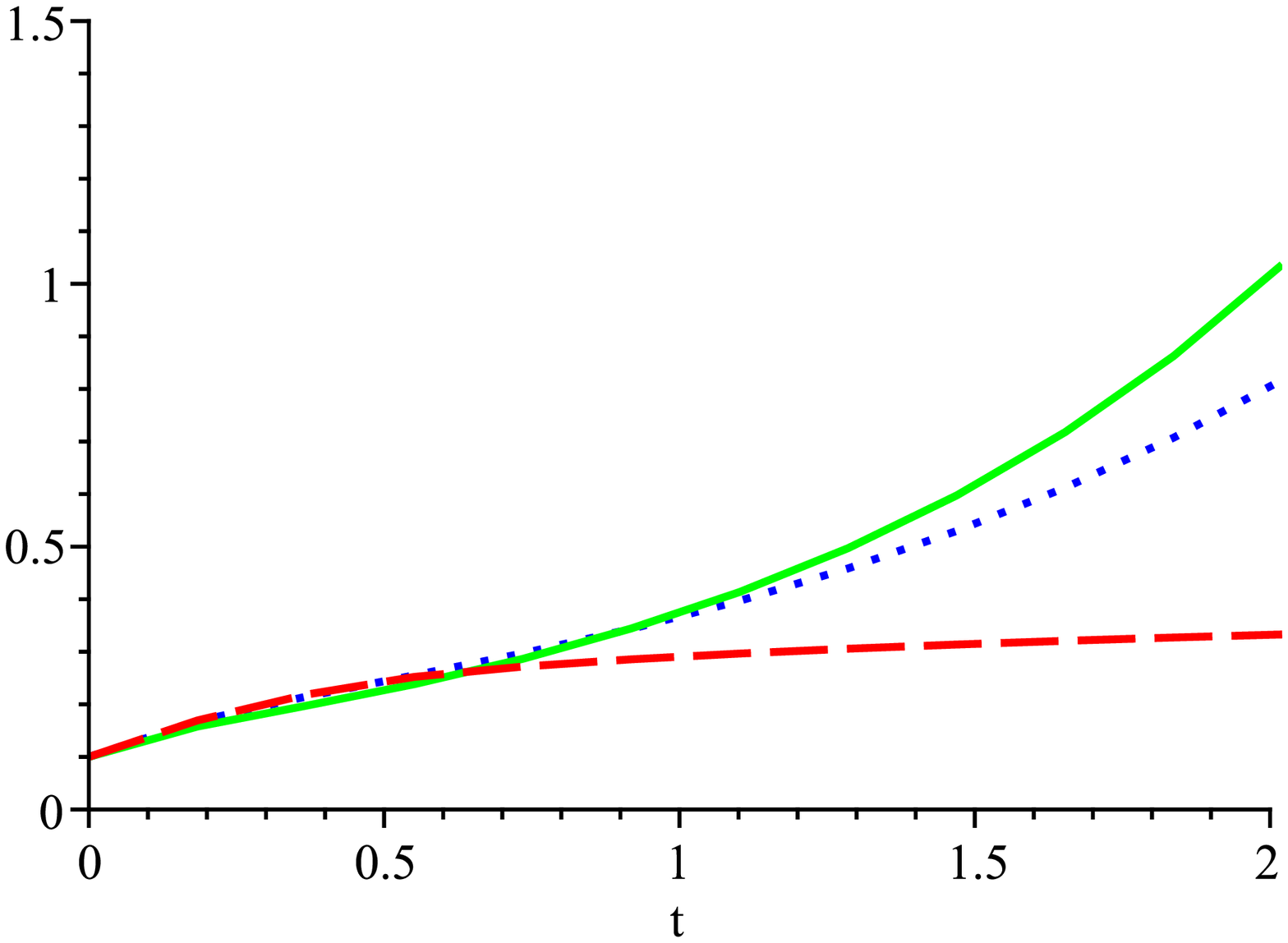}
		\caption{The behavior of $R(t)$ in 7-dimensions for $w=-\frac{3}{4}$ (left panel) and $w=1$ (right panel) and for $\alpha=1$ and $\beta= -1, 1, 0$ from up to down, respectively.}\label{figrtvet}
	\end{center}
\end{figure}

\begin{figure}
	\begin{center}
		\includegraphics[scale=0.35]{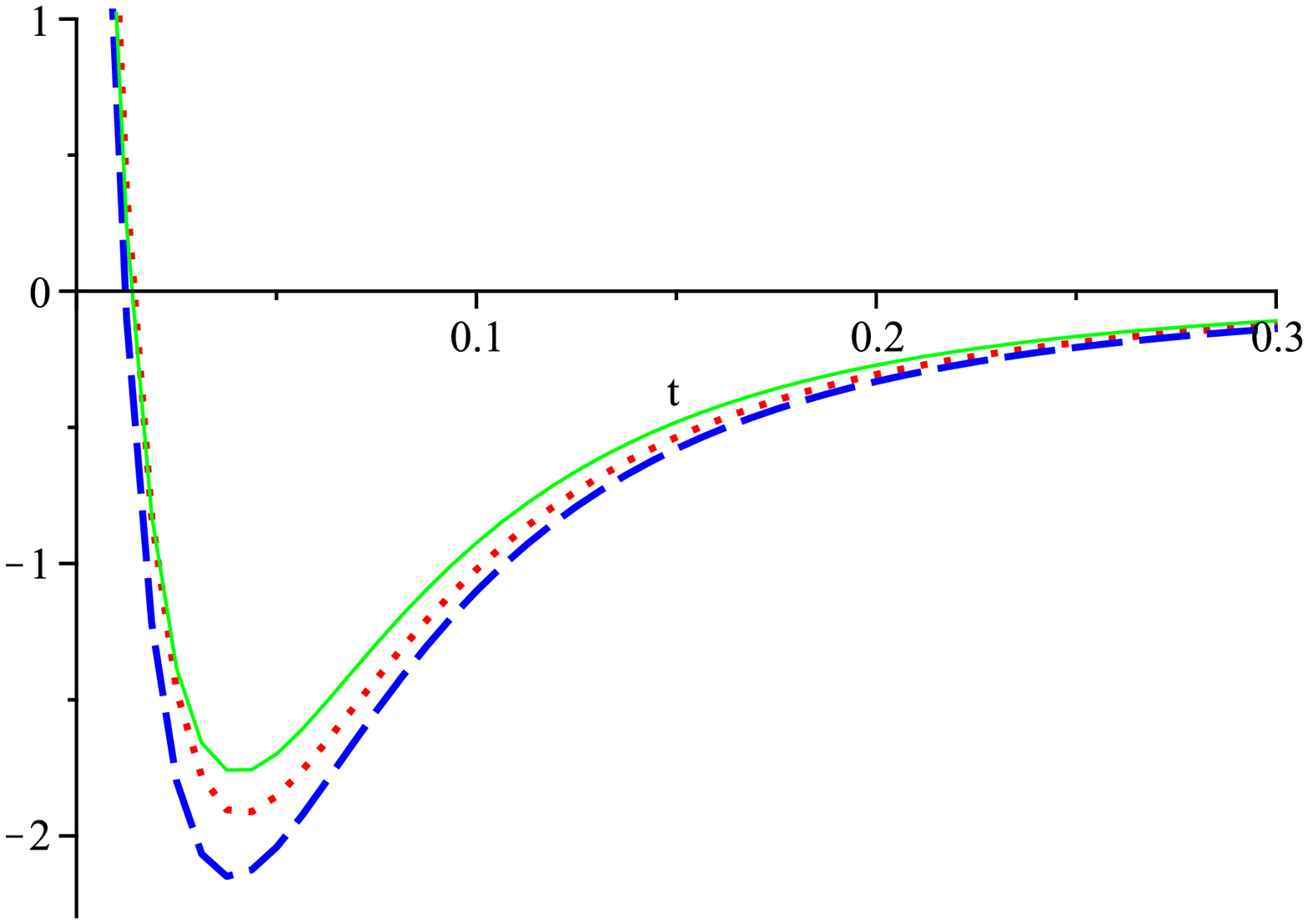}
		\includegraphics[scale=0.35]{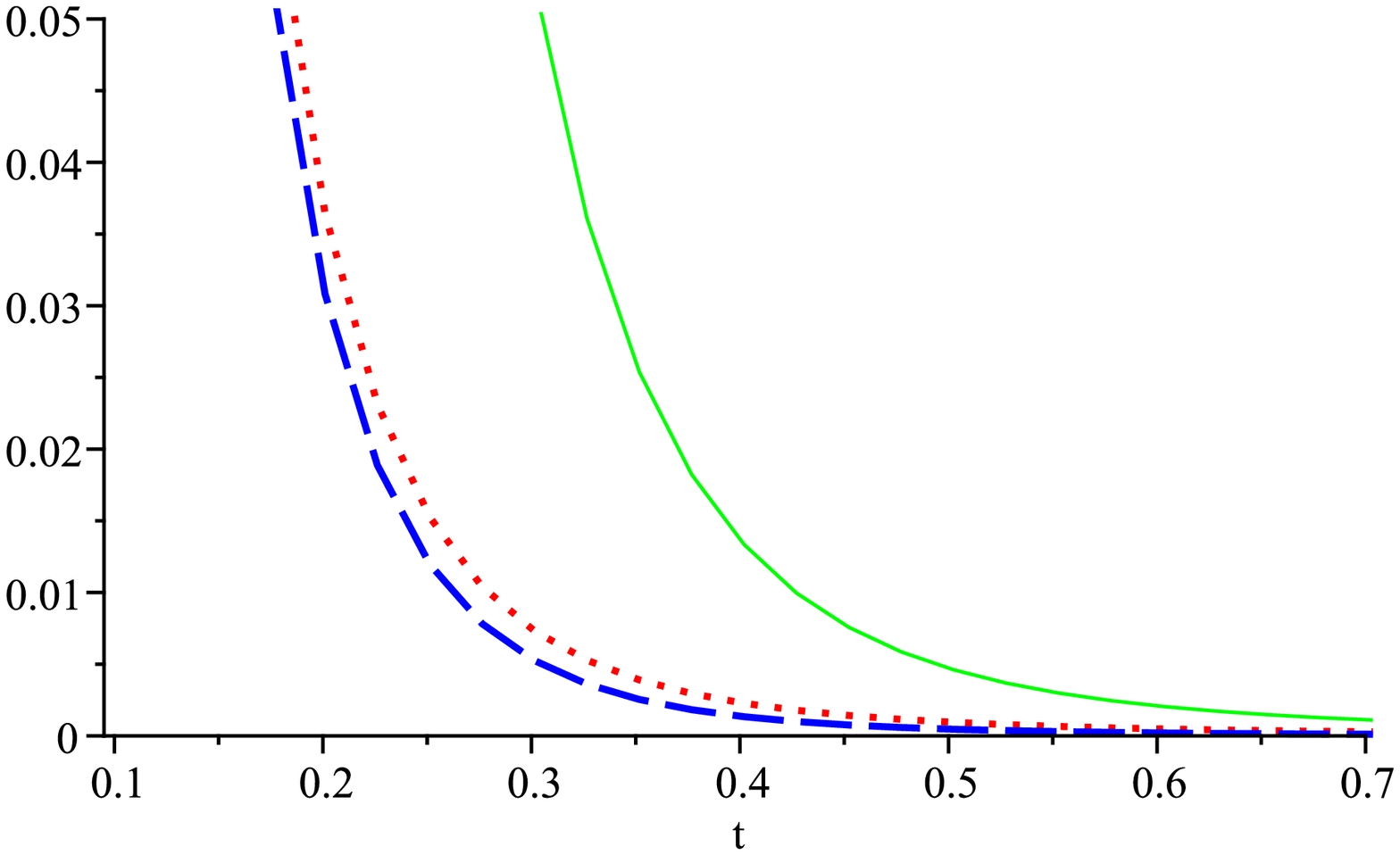}
		\caption{The behavior of $\protect\rho +p_{t}$, $\rho$, and $\protect\rho +p_{r}$ with respect to time at the throat $r_0=3$ for the left panel and at ${r=20}$ for the right panel, from top to bottom, respectively. The model parameters are chosen as $\alpha=1$, $\beta=1$ and $w=1$ in seven dimensions.}\label{figw1c0}
	\end{center}
\end{figure}

\begin{figure}
	\begin{center}
		\includegraphics[scale=0.35]{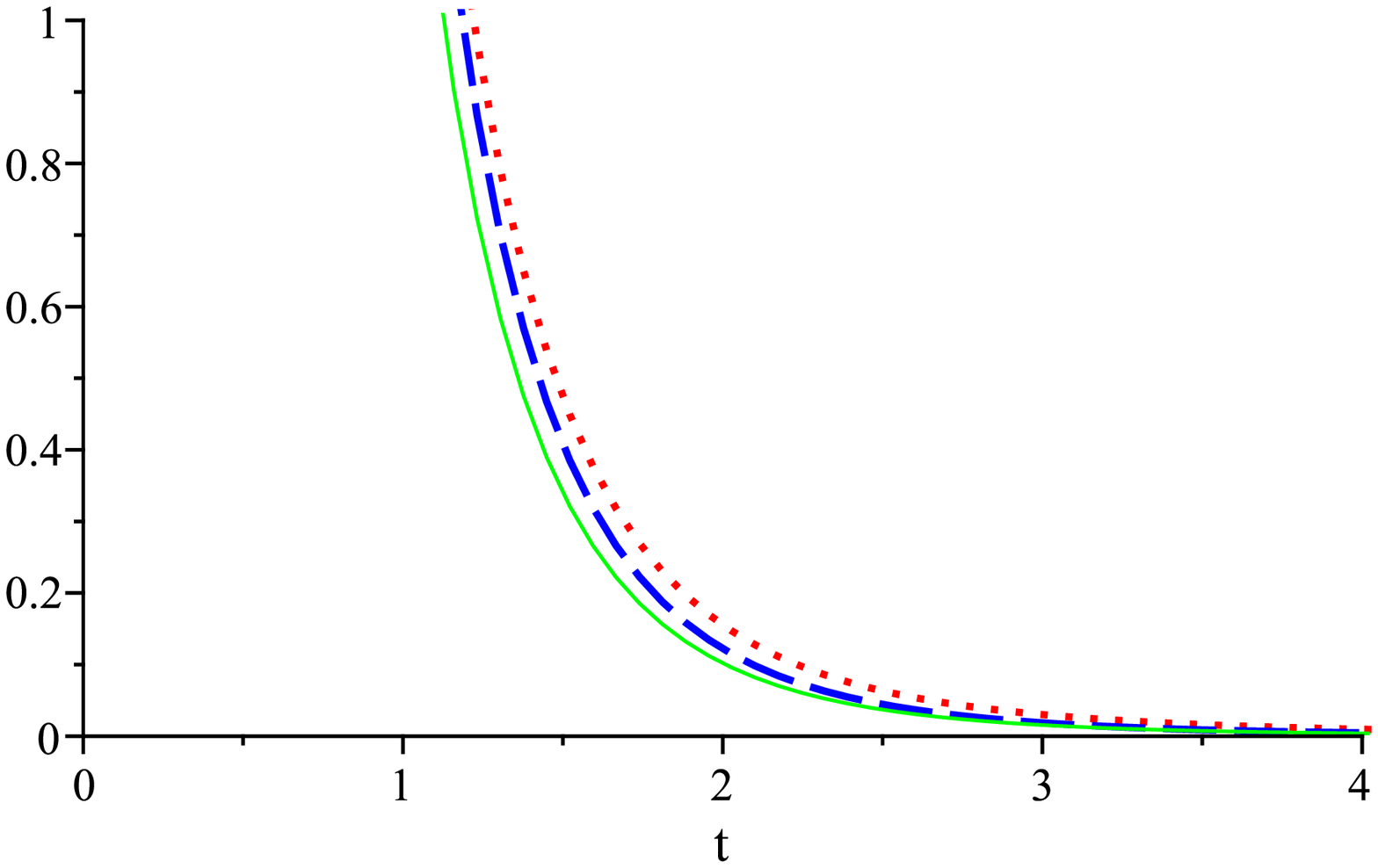}
		\includegraphics[scale=0.35]{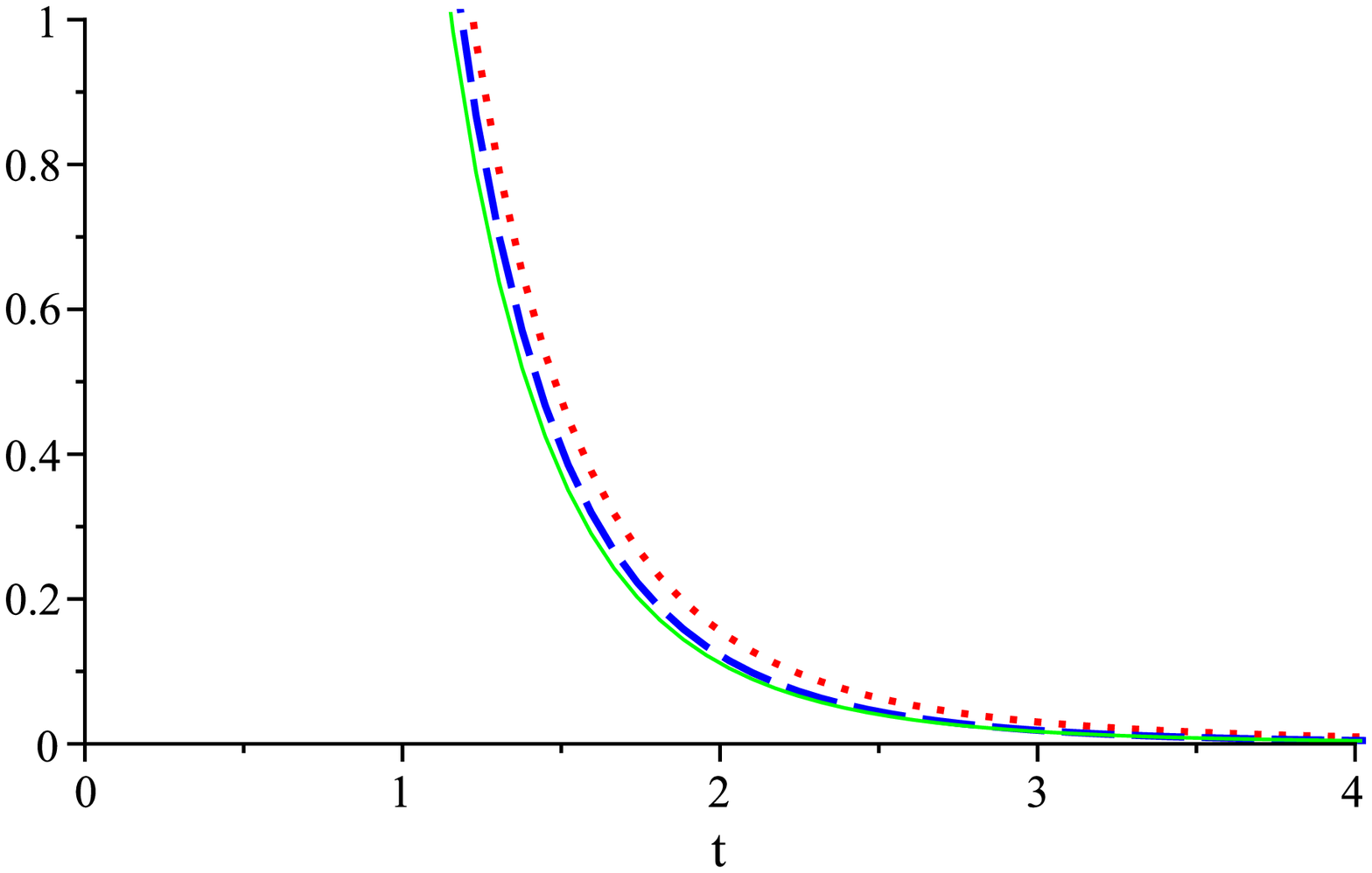}
		\caption{The behavior of $\rho$, $\protect\rho +p_{r}$, and $\protect\rho +p_{t}$ with respect to time at the throat $r_0=3$ for the left panel and at ${r=30}$ for the right panel from top to bottom, respectively. The model parameters are chosen as $\alpha=1$, $\beta=1$ and $w=-\frac{3}{4}$ in seven dimensions.}\label{figw34c0}
	\end{center}
\end{figure}

Since Eq.~(\ref{dynamic1}) cannot be solved analytically for $R(t)$, we proceed to numerically integrating this equation for a few values of model parameters, $\alpha$, $\beta$ and $w$, and investigate the WEC. For this purpose, we can substitute $C_1=0$ into Eq.~(\ref{dynamic1}) and obtain 
\begin{align}
{\frac {d}{dt}}H(t)+{\frac {H(t) ^{2} \left( 1+\alpha H(t)^{2}+\beta H(t)^{4}\right)\left(w+1\right)\left(n-1\right) }{2w+4w\alpha H (t) ^{2}+6w\beta H(t)^{4}}}=0.
\end{align} 
Now, in order to check the WEC, we first investigate the behavior of $\rho(r)$ and $\rho +p_{r}$ for large $r$, which is given by the following
approximations:
\begin{equation}
\rho\simeq \frac{(n-2)(n-1)H(t)^{2}}{2} \left[ 1+\alpha H(t)^{2}+\beta H(t)^4\right]+{\mathcal O}\left(\frac{1}{r^{2n-2}}\right),
\end{equation}
and
\begin{equation}
\rho +p_{r}=\rho +p_{t}\simeq \frac{(n-2)(n-1)(w+1)H(t)^2}{2} \left[1+\alpha H(t)^2+\beta H(t)^4 \right]+{\mathcal O}\left(\frac{1}{r^{n-1}}\right).
\end{equation}%
It is seen that for $w>-1$, $\alpha>0$, and $\beta>0$, both $\rho$ and $\rho+p_r$ are positive in the limit of large $r$ and consequently
the WEC is satisfied. In Fig.~(\ref{figrtvet}) the scale factor versus time is plotted for $\beta=-1,0,1$, $\alpha=1$  in seven dimensions and for $w=-\frac{3}{4}$ (left panel) and $w=1$ (right panel). Using then the field equations $(\ref{feild1})-(\ref{feild3})$ {along with} numerical values of the scale factor, {we can estimate the behavior of expressions for the WEC}. The numerical results are plotted in Figs.~(\ref{figw1c0}) and (\ref{figw34c0}). In these figures we can choose suitable values for the model parameters so that the WEC will be satisfied at the throat of the wormhole. In Fig.~(\ref{figw1c0}) we choose the parameters to be $\alpha=1$, $\beta=1$, and $w=1$ with $r_0=3$ in seven dimensions. We see that for large time the WEC is violated at the throat ({left panel}), {however, this condition holds as we move a way from the wormhole: see the right panel}. In Fig.~(\ref{figw34c0}), we depict the quantities $\rho$, $\rho+P_r$, and $\rho+P_t$ at the throat and {at larger radial distances} for positive values of the $\beta$ and $\alpha$ parameters and $w=-\frac{3}{4}$, where we observe that all of these quantities are satisfied. Hence, {for these parameter values the WEC is satisfied at all times and for $r\geq r_0$.} {We further note that, as we observe from Eq.~(\ref{rhoprr0}), static wormhole solutions for positive Gauss-Bonnet and Lovelock coefficients always violate energy conditions due to the flare-out condition. Such a violation of energy conditions holds in static wormhole configurations, for all positive coefficients of higher-order Lovelock gravity. While for dynamic wormhole configurations, e.g., the cases with $C_1=0$ and $w>-1$, one can pick out suitable positive values of Gauss-Bonnet and Lovelock coefficients in order to fulfill the energy conditions at the throat, see Figs. (\ref{figw1c0}) and (\ref{figw34c0}). In comparison to dynamic wormholes in Gauss-Bonnet gravity, for which negative Gauss-Bonnet coefficients lead to the satisfaction of energy conditions throughout the spacetime, the presence of the third-order Lovelock term helps to meet energy conditions for positive Gauss-Bonnet coefficients.}

\subsection{Solutions for the case $C_1=-1$}
\begin{figure}
	\begin{center}
	\hspace{-2.4cm}	\includegraphics[scale=0.43]{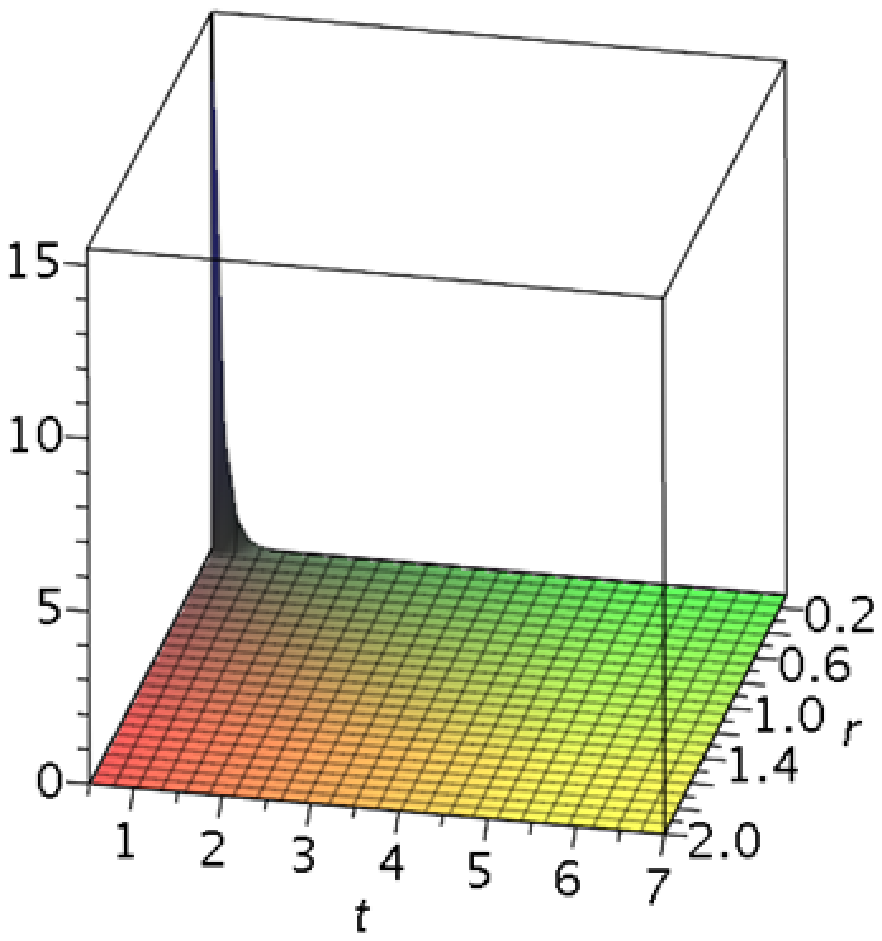}\hspace{-2.0cm}
		\includegraphics[scale=0.43]{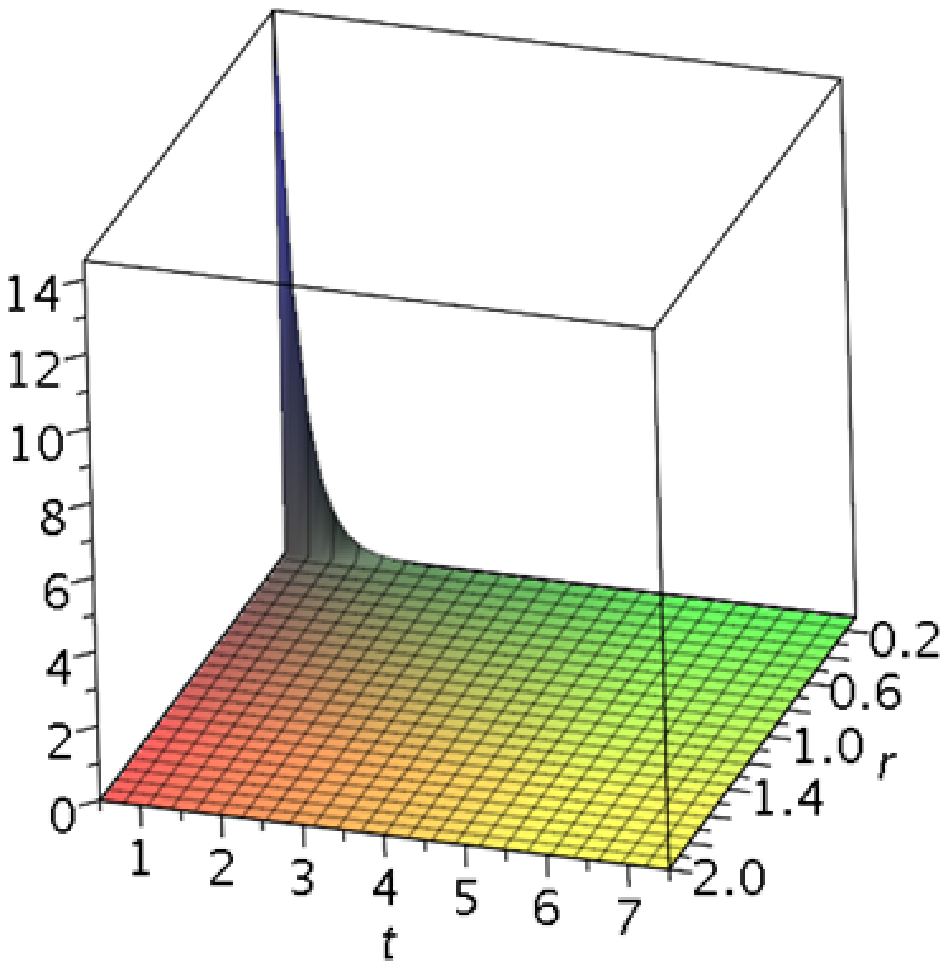}\hspace{-1.9cm}
		\includegraphics[scale=0.43]{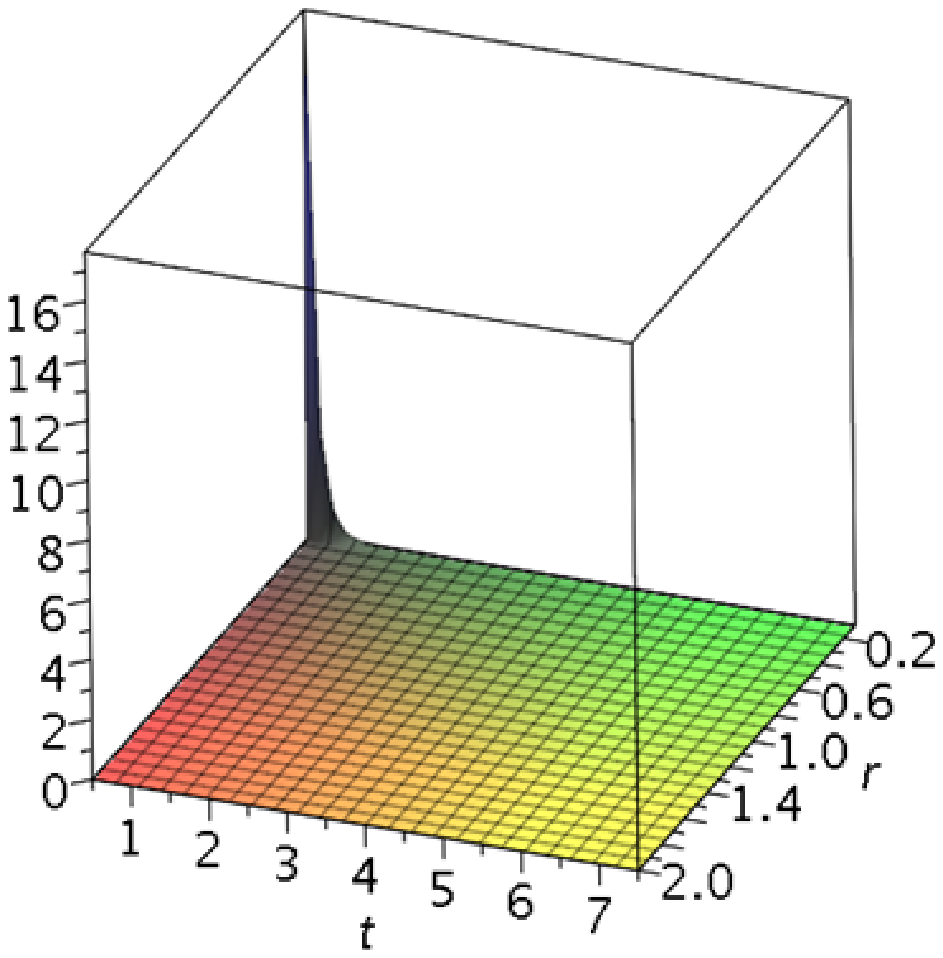}\hspace{-2cm}\vspace{-2.5cm}
		\caption {The behavior of $\rho$, $\rho +p_{r}$, and $\rho +p_{t}$ versus $r$ and $t$ respectively from left to right, for $w=-1$, $r_0=0.1$, $\alpha=1$, $\beta=-1$ and ${R_3=1.5}$ in seven dimensions.}\label{fg1}
	\end{center}
\end{figure}

In this subsection, we study the open background by using Eq.~(\ref{dynamic1}). We then find the analytical scale factor for $w=-1$ as 
\begin{align}
R \left( t \right) ={R_3}\,\sinh \left( {\frac {t}{{R_3}}}
\right),
\end{align}
where $R_3$ is a positive constant. In this case, we obtain the quantities $\rho$, $\rho+P_r$, and $\rho+P_t$ as
\begin{eqnarray}
\rho(r,t)&=&\frac{\left( n-1 \right)\left( n-2 \right) }{2R_3^{2}}+\notag \\
	&&\alpha\left({\frac { \left( n-1 \right)  \left( n-2 \right) }{2R_3^{4}}
	}-\frac{\left( n-1 \right) \left( {r_0}^{2}+1 \right) ^{2} \left( n-2 \right) 
	}{2R_3^{4}r_0^{4}\sinh^4\left({\frac {t}{R_3}} \right)}\left(\frac{r_0}{r}\right)^{2n-2}\right)+\notag \\
	&&\beta\left({\frac { \left( n-2 \right)  \left( n-1 \right) }{2R_3^{6}}
	}-\frac{3\left( n-2 \right) \left( n-1 \right)\left( r_0^{2}+1 \right)^{2}}{2{r_0}^4R_3^{6}\sinh^4 \left( {\frac{t}{R_3}}\right)}\left(\frac{r_0}{r}\right)^{2n-2}\right)-\notag\\&&\frac{\beta\left( n-2 \right)  \left( n-1 \right)\left( {{r_0}}^{2}+1 \right)^{3} }{R_3^{6}r_0^{6} \sinh^6 \left({\frac{t}{{R_3}}}\right) 
	}\left(\frac{r_0}{r}\right)^{3n-3}
\label{openwe1}
\end{eqnarray}

\begin{eqnarray}
\rho+P_r&=& -\frac{\left(r_0^{2}+1 \right)  \left( n-2 \right)  \left( n-3 \right)}{2r_0^{2} R_3^{2}\sinh^{2}\left({\frac {t}{R_3}} \right)	}\left(\frac{r_0}{r}\right)^{n-1}+\notag \\&&\alpha\left(-\frac{\left( _0^{2}+1 \right)\left( n-2 \right)\left( n-3 \right) }{r_0^{2}R_3^{4}\sinh^{2}\left( {\frac {t}{R_3}}\right)}\left(\frac{r_0}{r}\right)^{n-1}-\frac{\left(r_0^{2}+1 \right) ^{2} \left( n-2 \right)  \left( n-3 \right)}{R_3^{4}r_0^{4} \sinh^{4}\left( {\frac {t}{R_3}}\right)}\left(\frac{r_0}{r}\right)^{2n-2}\right)-\notag \\&&\beta\left(\frac{3\left( n-2 \right)\left( n-3 \right)\left(r_0^{2}+1 \right)}{2R_3^{6}\sinh^{2}\left(\frac{t}{R_3} \right)r_0^2
			}\left(\frac{r_0}{r}\right)^{n-1}+\frac{3\left( n-2 \right)\left( n-3 \right)\left( r_0^{2}+1 \right)^2}{R_3^{6}\sinh^{4}\left(\frac {t}{R_3}\right)r_0^4}\left(\frac{r_0}{r}\right)^{2n-2}\right)-\notag \\&&\frac{3\beta\left( n-2\right)  \left( n-3 \right)  \left( r_0^{2}+1 \right)^3}{2R_3^{6} \sinh^{2}\left(\frac {t}{R_3}\right)r_0^6}\left(\frac{r_0}{r}\right)^{3n-3}
			\label{openwe2}
\end{eqnarray}

\begin{eqnarray}
\rho+P_t&=&\frac{\left( r_0^{2}+1 \right)\left( n-3 \right) }{2r_0^{2}R_3^{2}\sinh^{2}\left({\frac{t}{R_3}}
	\right)}\left(\frac{r_0}{r}\right)^{n-1} +\notag \\&&\alpha\,\left(\frac{\left( r_0^{2}+1 \right)  \left( n-3 \right)}{r_0^{2}R_3^{4} \sinh^{2}\left({\frac{t}{R_3}}\right)}\left(\frac{r_0}{r}\right)^{n-1}-\frac{\left(r_0^{2}+1 \right) ^{2} \left( n+1 \right)  
}{R_3^{4}{{r_0}}^{4}\sinh^4\left({\frac {t}{{R_3}}}\right)}\left(\frac{r_0}{r}\right)^{2n-2}\right) +\notag \\&&\beta\left(\frac{3\left( r_0^{2}+1 \right)\left( n-3 \right) }{2R_3^{6}r_0^2\sinh^2\left({\frac {t}{R_3}}\right)}\left(\frac{r_0}{r}\right)^{n-1}-\frac{3\left( r_0^{2}+1 \right)^2 \left(n+1\right) }{R_3^{6}r_0^4\sinh^4\left( {\frac{t}{R_3}} \right)}\left(\frac{r_0}{r}\right)^{2n-2}\right)-\notag \\&&\frac{3\beta\left(r_0^{2}+1 \right)^3  \left( 3n-1 \right) }{2R_3^{6} r_0^6\sinh^6\left({\frac{t}{R_3}}\right)}\left(\frac{r_0}{r}\right)^{3n-3}.
\label{openwe3}
\end{eqnarray}
It is seen that in GR ($\alpha=0$ and $\beta=0$ ) $\rho+P_r$ is always negative, implying the violation of NEC throughout the spacetime. Let us now obtain $\rho$, $\rho+P_r$, and $\rho+P_t$ at the throat of the wormhole for small times,
\begin{align}
	\rho(r_0)=-{\beta\frac{\left( n-1 \right)\left( n-2 \right)\left(r_0^{2}+1 \right)^{3}}{r_0^{6}t^6}}-\frac{\left( n-1 \right)\left(n-2\right)  \left[\left( \beta+\alpha R_3^{2} \right)r_0^{2}-2\beta \right]\left(r_0^{2}+1 \right)^{2}}{{2r_0^{6} R_3^{2}t^4}}+{\mathcal O}\left(\frac{1}{t^{2}}\right),
\label{op0}
\end{align}

\begin{align}
	\rho+P_r\Big|_{r=r_0}=-{\frac {3 \left( r_0^{2}+1 \right) ^{3} \left( n-2\right)  \left( n-3 \right) \beta}{2r_0^{6}t^6}}-{\frac { \left(r_0^{2}+1 \right) ^{2} \left( n-2\right)  \left( n-3 \right)  \left( 3\beta r_0^{2}-3\beta+2\alpha r_0^{2}R_3^{2} \right) }{2r_0^{6}R_3^{2}t^4}}+{\mathcal O}\left(\frac{1}{t^{2}}\right),
	\label{op1}
\end{align}
and
 \begin{align} 	\rho+P_t\Big|_{r=r_0}=-{\frac{3\left(3n-1\right)\left(r_0^{2}+1\right)^{3}\beta}{2r_0^{6}t^6}}-{\frac{\left(  r_0^{2}+1 \right)^{2} \left[\left(\left(9-3n \right)\beta+2\alpha R_3^{2} \left( 1+n \right)\right) r_0^{2}-3\beta\left(3n-1 \right) \right] }{2r_0^{6} R_3^{2}t^4}}+{\mathcal O}\left(\frac{1}{t^{2}}\right).	
\label{op2}
 \end{align}
One can easily show that {in the limit $t\rightarrow0$ the WEC is satisfied for $\beta<0$ and arbitrary values of the $\alpha$ parameter}. This is due to the presence of the first term in Eqs.~(\ref{op0})-(\ref{op2}). {One can then suitably choose the model parameters so that the WEC will be satisfied for $r>r_0$ and for all times}; see Fig.~(\ref{fg1}).

\subsection{Solutions for the case $C_1=1$}
\begin{figure}
	\begin{center}
		\includegraphics[scale=0.35]{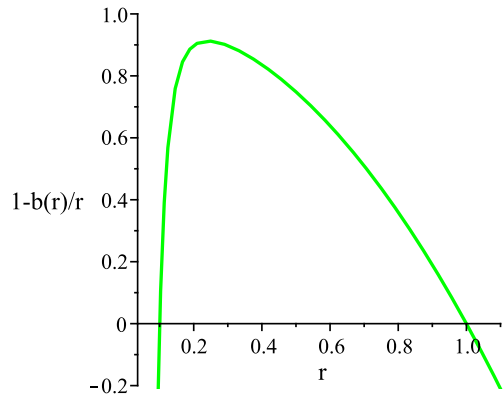}
		\caption{The behavior of $1-b(r)/r$ with respect to $r$ for $r_0=0.1$, $n=7$ with $b^{\prime}(r_0)=-2.94$.}\label{figbr}
	\end{center}
\end{figure}

\begin{figure}
	\begin{center}
		\includegraphics[scale=0.26]{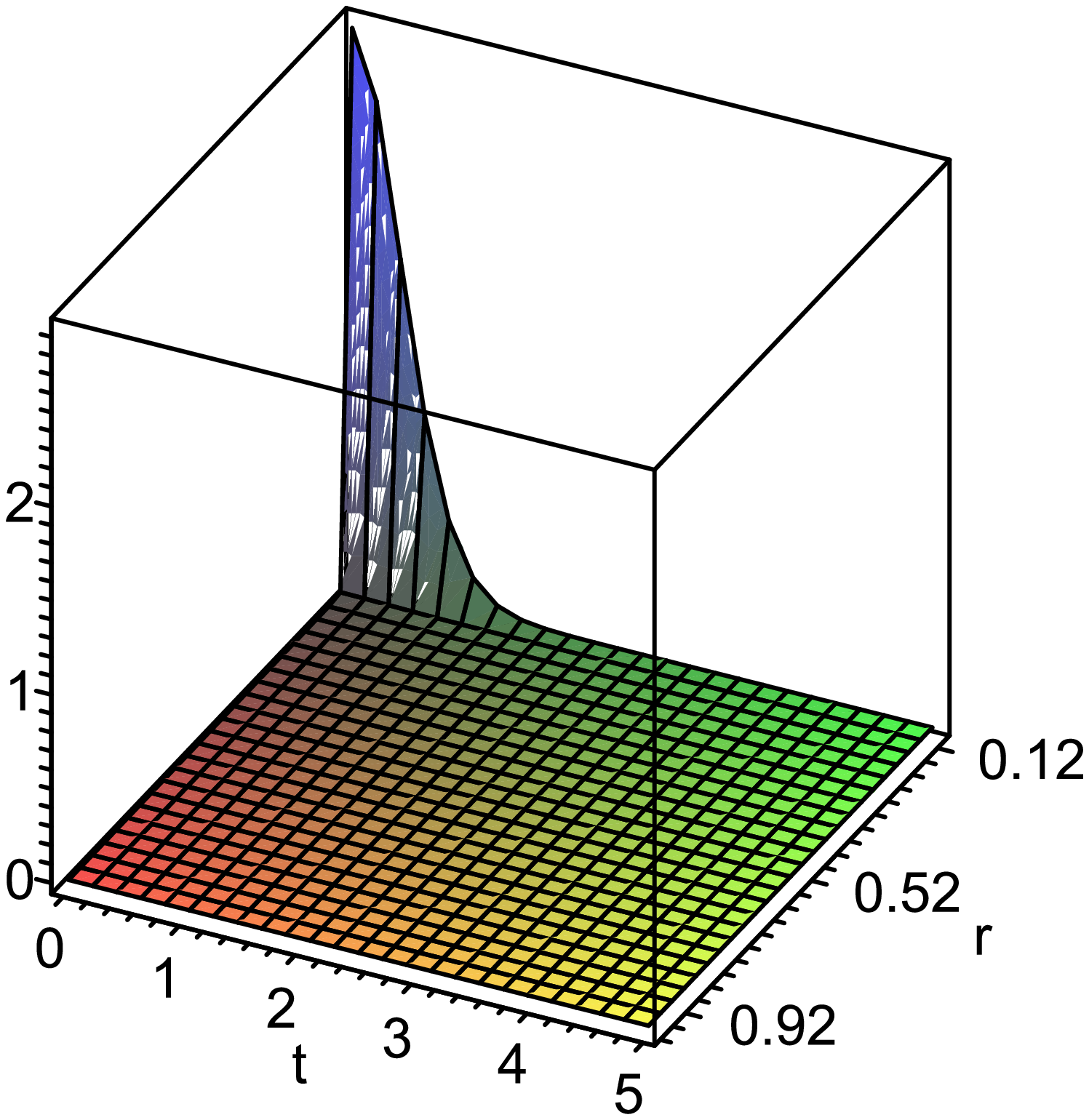}
		\includegraphics[scale=0.26]{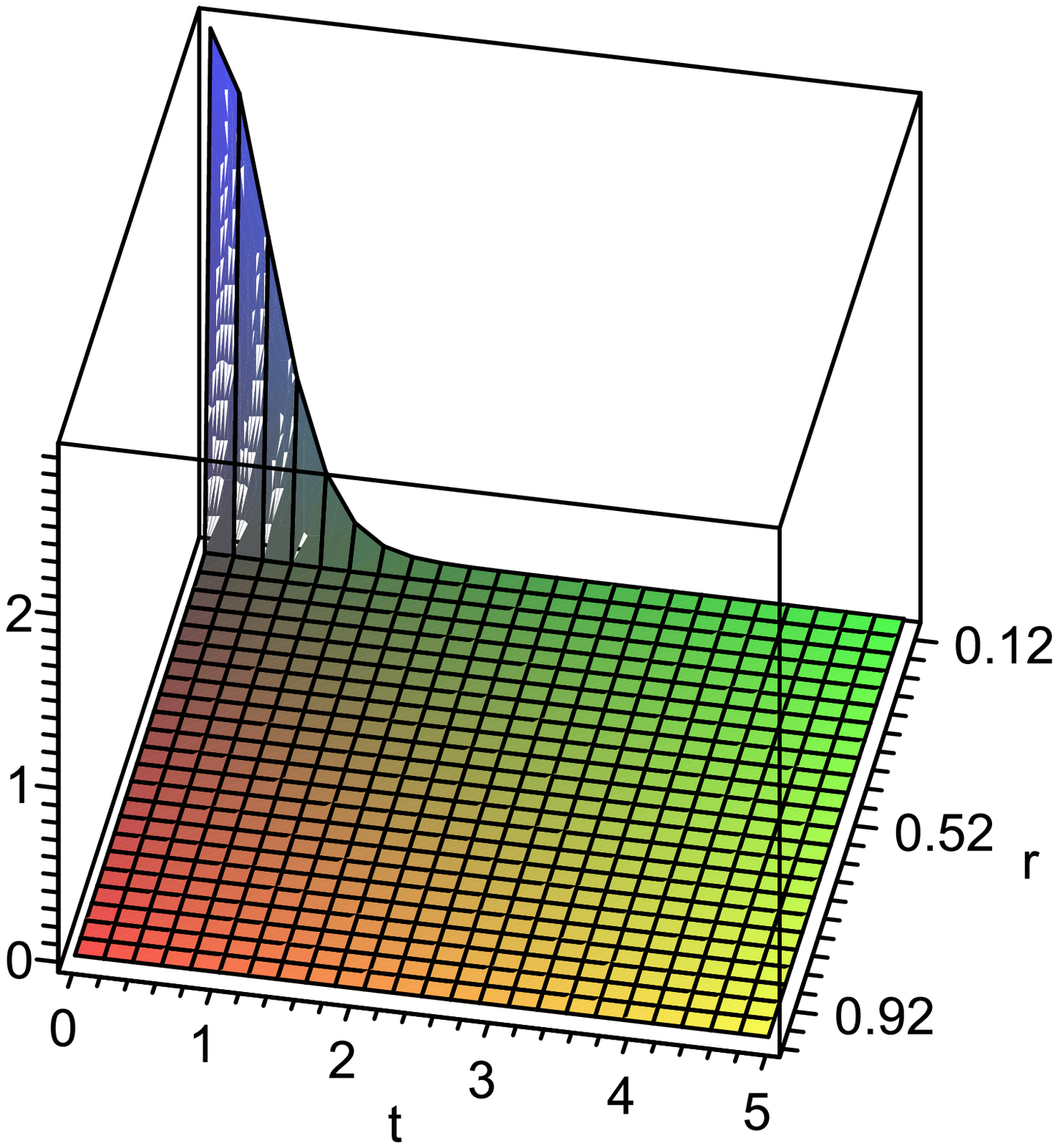}
		\includegraphics[scale=0.26]{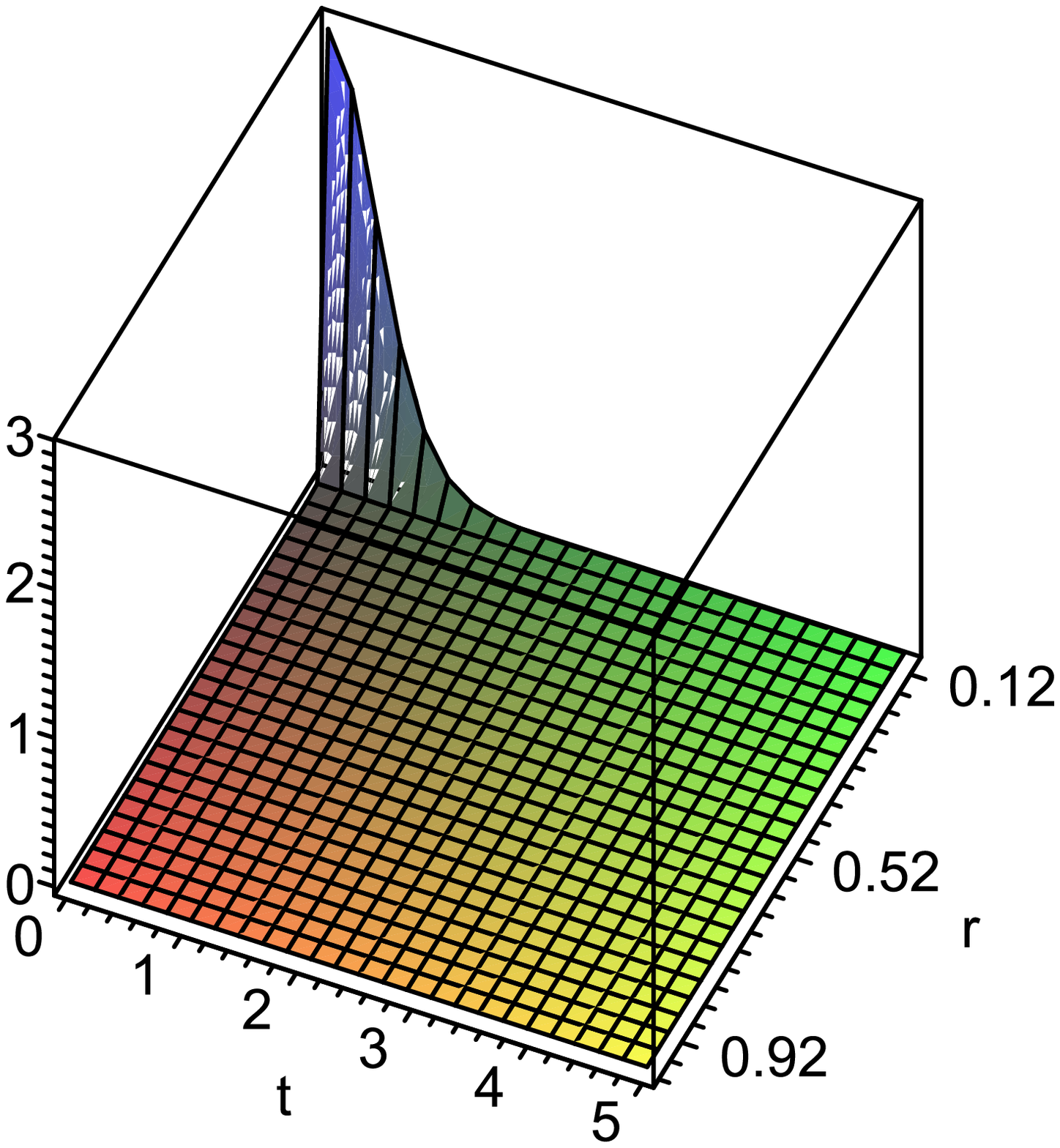}
		\caption {The behavior of $\rho$, $\rho +p_{r}$, and $\rho +p_{t}$ versus $r$ and $t$ respectively from left to right, for $w=-1$, $r_0=0.1$, $R_4=1$, $\alpha=1$ and $\beta=-1$ in  seven dimensions.}\label{fg44}
	\end{center}
\end{figure}

In the case of a closed background, we can choose the wormhole throat such that the condition $b^{\prime}(r_0)<1$ is satisfied, i.e.,  $r_0^2<\frac{n-3}{n-1}$. To be a solution of a wormhole, the condition $0 < r- b(r)$ is also imposed. The condition $b(r) = r$ leads to two real and positive roots given by $r_{-} = r_0$ and $r_{+}$, which is given by
\begin{align}
r_{+}^{n-1}-r_{+}^{n-3}-r_0^{n-1}+r_0^{n-3}=0.\label{ss2}
\end{align}
Thus, the spatial extension of this type of wormhole solution cannot be arbitrarily large. We then have a finite wormhole configuration within the range $r_{-}<r<r_{+}$. For instance, consider $n=7$ in Eq.~(\ref{ss2}). The $r_{+}$ is then found as
\begin{align}
r_+=\frac{\sqrt {2-2r_0^{2}+2\sqrt {1+2r_0^{2}-3r_0^{4}}}}{2}.
\end{align} 
In Fig.~(\ref{figbr}), we plot the quantity $1-b(r)/r$ versus $r$ for $n=7$ and as it is observed the condition $b^{\prime}(r_0)<1$ is satisfied at the throat. Also, increasing the dimension of space enlarges the wormhole spatial extension. In order to study energy conditions for this class of solutions we proceed with obtaining the behavior of the scale factor, using Eq. (\ref{dynamic1}) for $w=-1$, as
 \begin{align}
R\left(t\right)=R_4\cosh\left({\frac{t}{R_4}}\right),
 \end{align} 
where $R_4$ is a constant. In order to check the WEC we can substitute the above solution into the expressions (\ref{feild1})-(\ref{feild3}) to obtain
\begin{eqnarray}
\rho(r,t)&=&\frac { \left( n-1 \right)  \left( n-2 \right) }{2R_4^{2}}+\notag \\
	&&\alpha\,\left({\frac { \left( n-1 \right)  \left( n-2 \right) }{2 R_4^{4}}
	}-\frac{\left(n-1\right) \left(r_0^{2}-1\right)^{2} \left( n-2 \right)}{2R_4^4r_0^{4}\cosh^4
	\left( {\frac {t}{R_4}} \right)}\left(\frac{r_0}{r}\right)^{2n-2}	\right)+\notag \\
&&\beta\left({\frac { \left( n-2 \right)  \left( n-1 \right) }{2R_4^{6}}}-\frac{3 \left( n-2 \right)  \left( n-1 \right)  \left(r_0^{2}-1
	\right) ^{4}}{2r_0^4R_4^{6}\cosh^4\left( {\frac {t}{R_4}} \right)
	}\left(\frac{r_0}{r}\right)^{2n-2}\right)+\notag\\&&\beta\left(\frac{\left( n-2 \right)  \left( n-1 \right)\left(r_0^{2}-1
		\right)^{6}}{r_0^6R_4^{6}\cosh^{6}\left( {\frac{t}{R_4}} \right)}\left(\frac{r_0}{r}\right)^{3n-3}\right)
\label{closewe1}
\end{eqnarray} 
\begin{eqnarray}
\rho+P_r&=& \frac{\left(r_0^{2}-1 \right)\left( n-2 \right)\left( n-3 \right)}{2r_0^{2}R_4^{2}\cosh^2\left({\frac{t}{R_4}} \right) }\left(\frac{r_0}{r}\right)^{n-1}+\notag \\&&\alpha\left(\frac{\left(r_0^{2}-1 \right)\left( n-2 \right)  \left( n-3 \right)}{r_0^{2}R_4^{4}\cosh^2\left( {\frac{t}{R_4}}\right)}\left(\frac{r_0}{r}\right)^{n-1}-\frac{\left(r_0^{2}-1 \right) ^{2} \left( n-2 \right)\left( n-3 \right)}{R_4^{4}r_0^{4}\cosh^4\left( {\frac {t}{R_4}}
\right)}\left(\frac{r_0}{r}\right)^{2n-2}\right)+\notag \\&&\beta\left(\frac{3\left(r_0^{2}-1 \right)\left( n-2 \right)\left( n-3 \right) 
}{2r_0^{2}R_4^{6}\cosh^2\left( {\frac{t}{R_4}}\right)}\left(\frac{r_0}{r}\right)^{n-1}-\frac{3\left(r_0^{2}-1 \right)^{4} \left( n-2 \right)  \left( n-3 \right)}{R_4^{6}r_0^{4}\cosh^4\left( {\frac {t}{R_4}}\right)
}\left(\frac{r_0}{r}\right)^{2n-2}\right)+\notag\\&&\beta\left(\frac{3\left(r_0^{2}-1 \right) ^{6} \left( n-2 \right)  \left( n-3 \right) 
}{2R_4^{6}r_0^{6}\cosh^{6}\left( {\frac{t}{R_4}}\right)
}\left(\frac{r_0}{r}\right)^{3n-3}\right)
\label{closewe2}
\end{eqnarray}
\begin{eqnarray}
\rho+P_t&=&-\frac{\left(r_0^{2}-1 \right)\left( n-3 \right)}{2r_0^{2}R_4^{2} \cosh^{2}\left({\frac{t}{R_4}}\right)}\left(\frac{r_0}{r}\right)^{n-1} -\notag \\&&\alpha\left(\frac{\left(r_0^{2}-1\right)\left(n-3\right)}{r_0^{2}R_4^{4}\cosh^{2}\left({\frac {t}{R_4}}\right)}\left(\frac{r_0}{r}\right)^{n-1}+\frac{\left(r_0^{2}-1\right)^{2}\left( n+1 \right)  
}{R_4^{4}r_0^{4}\cosh^4\left( {\frac {t}{R_4}}\right)
}\left(\frac{r_0}{r}\right)^{2n-2}\right)-\notag \\&&\beta\left(\frac{3\left(r_0^{2}-1 \right)  \left( n-3 \right) 
}{2r_0^{2}R_4^{6}\cosh^2\left( {\frac {t}{R_4}}
\right)}\left(\frac{r_0}{r}\right)^{n-1}+\frac{3\left(r_0^{2}-1 \right)^{4}\left( n+1 \right) 
}{R_4^{6}r_0^{4}\cosh^{4}\left( {\frac {t}{R_4}}
\right)}\left(\frac{r_0}{r}\right)^{2n-2}\right)+\notag\\&&\beta\left(\frac{3\left(r_0^{2}-1 \right)^{6}\left( 3n-1 \right) 
}{2R_4^{6}r_0^{6}\cosh^6\left( {\frac {t}{R_4}}
\right)}\left(\frac{r_0}{r}\right)^{3n-3}\right).
\label{closewe3}
\end{eqnarray}
Let us now obtain $\rho$, $\rho+P_r$, and $\rho+P_t$ at the throat the wormhole for small times,
\begin{eqnarray}
\rho(r_0)&=&{\frac { \left( n-1 \right)  \left(r_0^{6}R_4^{4}+2\alpha r_0^{4}R_4^{2}+ \left( 3\beta-\alpha R_4^{2} \right)r_0^{2}-2\beta \right)\left(n-2\right) }{2r_0^{6}R_4^{6}}}+\notag\\&&{\frac{\left( n-2 \right)  \left( n-1 \right)  \left(r_0^2-1
		\right)^{2}\left( 3\beta+\alpha r_0^{2}R_4^{2} \right)}{r_0^{6}R_4^{8}}}t^2+ {\mathcal O}\left(t^4\right)
\label{occ1}
\end{eqnarray}
\begin{eqnarray}
\rho+P_r\Big|_{r=r_0}\!\!\!\!\!\!\!\!\!\!\!\!\!\!&=&{\frac { \left( n-2 \right)  \left( n-3 \right) \left( {r_0^2}-1 \right)\left( 3\beta+2r_0^{2
		}\alpha R_4^{2}+r_0^{4}R_4^{4} \right) }{2r_0^{6}R_4^{6}}}-\notag\\&&{\frac{\left( n-2\right)\left(n-3\right)\left(r_0^2-1 \right)  \left(R_4^{2} \left(-2\alpha+R_4^{2} \right) r_0^{4}+ \left( -6\beta+4\alpha R_4^{2} \right)r_0^{2}+9\beta \right)}{2			r_0^{6}R_4^{8}}}t^2+{\mathcal O}\left(t^4\right),
\label{occ2}
\end{eqnarray}

\begin{eqnarray}
\rho+P_t\Big|_{r=r_0}\!\!\!\!\!\!\!\!\!\!\!\!&=&-\frac{\left(r_0^2-1\right)\left[\left( n-3 \right)r_0^{4}R_4^{4}+\left(\left(4n-4 \right) r_0^{4}-2\left( n+1 \right)R_4^{2}r_0^{2}\right)\alpha+\Sigma_1\right]}{{2r_0}^{6}{R_4^6}}+\notag\\&&\frac{\left(r_0^2-1\right)\left[\left(  \left( 6n-2 \right)R_4^{2}r_0^{4}-4\left( n+1 \right)R_4^{2}r_0^{2}\right)\alpha+\left( n-3 \right)r_0^{4}R_4^{4}+\Sigma_2\right]}{{2 r_0}^{6}{R_4^8}}t^2+{\mathcal O}\left(t^4\right),
\label{occ3}
\end{eqnarray}
where
\begin{align*}
\Sigma_1=\left[\left(12n-12\right)r_0^{2}-9n+3 \right]\beta,~
\Sigma_2=\left[\left(-12n+12\right)r_0^{4}+\left(42n-30 \right)r_0^{2}-27n+9 \right]\beta.
\end{align*}
Also, the {asymptotic} behavior of $\rho+P_t$ and $\rho+P_r$ is obtained as
\begin{eqnarray}
\rho+P_r\simeq\frac{\left( n-3 \right)  \left( n-2 \right)\left(r_0^{2}-1\right) r_0^{n-3} \left( 3\beta+R_4^{4}+2\alpha R_4^{2}
\right)}{2R_4^{6}r^{n-1}\cosh^2\left({\frac{t}{R4}}\right)}+{\mathcal O}\left(\frac{1}{r^{2n-2}}\right)
\end{eqnarray}
and
\begin{eqnarray}
\rho+P_t\simeq-\frac{\left(n-3 \right)\left(r_0^{2}-1\right)r_0^{n-3}\left(3\beta+R_4^{4}+2\alpha R_4^{2}
\right)}{2 R_4^{6}r^{n-1}\cosh^2\left({\frac{t}{{R4}}}\right)}+{\mathcal O}\left(\frac{1}{r^{2n-2}}\right).
\end{eqnarray}
In this case, we see that in GR {limit} ($\alpha=0$,~$\beta=0$), $\rho+P_t$ is always negative, implying the violation of NEC {at all times}. However, {for the present model,} we can choose suitable values for the $\alpha$ and $\beta$ parameters so that we have normal matter {for the wormhole configuration}. We note that, {asymptotically}, the behavior of $\rho+P_t$ and $\rho+P_r$ {is} in the opposite direction, then, one can {set} $\alpha=-{\frac {3\beta+R_4^{4}}{2R_4^{2}}}$ so that the WEC can {be} satisfied. In Figs.~(\ref{fg44}), we depict the quantities $\rho$, $\rho+P_r$, and $\rho+P_t$ in terms of $r$ and $t$ for $R_4=1, r_0=0.1$, and $n=7$.
\section{Concluding Remarks}\label{concluding}
In this paper, we have explored higher-dimensional dynamical wormhole solutions in the framework of Lovelock gravity by considering a constraint on the Ricci scalar. In this context, the existence of higher curvature terms may help to construct wormhole configurations that respect energy conditions. In a cosmological set up, microscopic dynamical wormholes produced in the early universe may be inflated to macroscopic scales. Our analysis shows that for all solutions with $w=-1$, one can choose the Lovelock coefficients as $\alpha>0$ and $\beta<0$ with a suitable value for the throat so that the wormhole solutions obtained in this manner respect the weak energy condition in whole space. For the case $C_1=0$, one can choose special values for second-order and third-order Lovelock coefficients so that the {WEC} will be respected at the throat. Also in this case, we can choose positive second-order and third-order Lovelock coefficients for $w>-1$ so that the WEC is {fulfilled asymptotically}. For $C_1=-1$ and suitable values for the wormhole throat, one can choose the Lovelock coefficients as $\alpha>0$ and $\beta<0$ in such a way that the energy conditions hold throughout the spacetime. Moreover, for $C_1=1$, $\alpha>0$, and $\beta<0$ the energy conditions are satisfied; however, wormhole configurations constructed in this way can exist within a small region of space. {It should be noted that the existence of a curvature singularity within the spacetime can be examined through investigating the behavior of the Kretschmann invariant (\ref{KRETC}) and others. In this sense, the divergence of the Kretschmann invariant at some spacetime event signals the occurrence of a spacetime singularity~\cite{KRSIN}. Regarding our solutions, we observe that this invariant behaves regularly for $r\geq r_0$ and thus, the spacetime geometry has no curvature singularity in this range.}
\par
As the final remarks concerning the future of research works, it is worth mentioning that during the past years several branches of theoretical physics such as string theory, supergravity and Kaluza-Klein theory have predicted the presence of extra dimensions~\cite{extradim}. It is therefore plausible to search for possible existence of geometrical compact objects within higher-dimensional spacetimes. For example higher-dimensional black holes, wormholes, and positive mass solutions with naked singularities~\cite{bhwhextra}. Moreover, the near horizon black hole solutions in higher-dimensional models are of particular interest since they can be regarded as windows to extra dimensions~\cite{davis}. From a cosmological perspective, the possible existence of extra dimensions is significant for inflationary scenarios~\cite{higherinf} and during the early cosmic times~\cite{duear}. In~\cite{pleu} the author has provided some observational criterion in order to determine whether the extra dimensions are compact or large and the phenomenological aspects of large, warped, and universal extra dimensions are reviewed in~\cite{extradim,extraphenom}. In this context, wormhole geometries without exotic matter have been studied in~\cite{karwo}. Such solutions could be thought of as similar to missing energies in collider phenomenology which are expected to provide signals of the existence of extra dimensions~\cite{randa}. Therefore, the existence of such configurations with extra dimensions in our universe cannot be a priori excluded, and their possible astrophysical results could be a subject of further investigations.
\par
\section*{Acknowledgements}{The authors would like to appreciate the anonymous referee for providing useful and constructive comments that helped us to improve the original version of our manuscript.} 

\end{document}